\documentclass[12pt,onecolumn,letterpaper]{IEEEtran}
\pdfoutput=1

\usepackage[pdftex]{graphicx}				
\usepackage{amssymb,amsmath}

\newtheorem{Thm}{Theorem}

\newtheorem{Lem}{Lemma}
\newtheorem{Cor}{Corollary}
\newtheorem{Def}{Definition}
\newtheorem{Fact}{Fact}

\title{Duality Codes and the Integrality Gap Bound for Index Coding}
\author{Hao Yu and Michael J. Neely\\University of Southern California
\thanks{The authors are with the  Electrical Engineering department at the University
of Southern California, Los Angeles, CA.} 
\thanks{This work is to be presented in part at the Allerton conference on communications, control, and computing, Monticello, IL, October, 2013. }
\thanks{This work is supported in part  by the NSF grant 0964479.}}

\begin{document}
\maketitle

\begin{abstract}
This paper considers a base station that delivers packets to multiple receivers through a sequence of coded transmissions.  All receivers overhear the same transmissions.  Each receiver may already have some of the packets as side information, and requests another subset of the packets.  This problem is known as the index coding problem and can be represented by a bipartite digraph.  An integer linear program is developed that provides a lower bound on the minimum number of transmissions required for any coding algorithm.  Conversely, its linear programming relaxation is shown to provide an upper bound that is achievable by a simple form of vector linear coding. Thus, the information theoretic optimum is bounded by the \emph{integrality gap} between the integer program and its linear relaxation.  In the special case when the digraph has a planar structure, the integrality gap is shown to be zero, so that exact optimality is achieved.  Finally, for non-planar problems, an enhanced integer program is constructed that provides a smaller integrality gap.  The dual of this problem corresponds to a more sophisticated \emph{partial clique coding strategy} that time-shares between Reed-Solomon erasure codes.  This work illuminates the relationship between index coding, duality, and integrality gaps between integer programs and their linear relaxations. 
\end{abstract}


\section{Introduction}

Consider a noiseless wireless system with $N$ receivers, $W$ independent packets of the same size, and a single broadcast station.  The broadcast station has all packets. Each receiver has a subset of the packets as side information, but desires another (disjoint) subset  of the packets.  The broadcast station must deliver the packets to their intended receivers.  To this end, it makes a sequence of (possibly coded) transmissions that are overheard by all receivers.   The goal is to find a coding scheme with the minimum number of transmissions (clearance time) such that each user is able to decode its demanded packets. This problem was introduced by Birk and Kol in \cite{Birk98INFOCOM,Birk06IT} and is known as the index coding problem.

The formulation of the index coding problem is simple, elegant and captures the essence of broadcasting with side information. It also relates directly to multi-hop network coding problems. Specifically, work in \cite{Rouayheb10IT} shows that an index coding problem can be reduced to a network coding problem. A partial converse of this result is also shown in \cite{Rouayheb10IT}, in that linear versions of network coding can be redued to linear index coding (see \cite{Effros12Arxiv} for extended results in this direction). However, the index coding problem still seems to be intractable. The first index coding problem investigated by Birk and Kol considers only the case of unicast packets and can be represented as a directed side information graph.  Work by Bar-Yossef et. al. in \cite{Yossef11IT} shows that the performance of the best \emph{scalar linear code} is equal to the graph parameter \emph{minrank} of the side information graph. However, computing the minrank of a given graph  is NP-hard \cite{Peeters96Combinatorica}.  Further, it is known that restricting to scalar linear codes is generally sub-optimal \cite{Alon08FOCS,Lubetzky09IT}.

One branch of research on index coding aims to find tight performance bounds. Work in \cite{Yossef11IT} shows that if the index coding problem has an \emph{undirected} side information graph (such as when it has symmetric demands) then the minrank is lower-bounded by the independence number of the graph, and upper-bounded by the clique cover number. For the unicast index coding problem, work in \cite{Yossef11IT} shows that the optimal clearance time (with respect to any scalar, vector or non-linear code) is lower-bounded by the maximum acyclic subgraph of the side information graph. Work in \cite{NeelyTehraniZhang12} generalizes this to the multi-cast case using a directed bipartite graph. It shows that the optimum of the general problem is lower-bounded by the maximum acyclic subgraph induced by deletions of packet vertices, user-vertices and packet-to-user arcs. In \cite{Blasiak10Arxiv}, a sequence of linear programs is proposed to bound the optimal clearance time.

Another branch of research on index coding focuses on studying the performance of specific codes and specific graph structures. Work in \cite{Alon08FOCS} shows that vector linear codes can have strictly better performance compared with scalar linear codes. Work in \cite{Lubetzky09IT} demonstrates that non-linear codes can outperform both scalar and vector linear codes.  Instead of finding the minimum clearance time, Chaudhry et. al. in \cite{Chaudhry11} consider the problem of maximizing the total number of saved transmissions by exploiting a specific code structure together with graph theory algorithms. Ong et. al. in \cite{Ong12ICC} find the optimal index code in the \emph{single uniprior} case, where each user only has a single uniprior packet as side information.

This paper studies index coding from a perspective of optimization and duality. It illustrates the inherent duality between the information theoretical lower bound in \cite{NeelyTehraniZhang12} and the performance of specific codes. Section \ref{section:weighted_bipartite_graph} extends the bipartite digraph representation of the problem to a weighted bipartite digraph. Section \ref{section:integer_program_lower_bound} uses this new graph structure to develop an integer linear program that finds the tightest lower bound given by \cite{NeelyTehraniZhang12}. Section \ref{section:cyclic-code-duality}  considers the linear programming (LP) relaxation of the integer program, and shows that the dual problem of this relaxation corresponds to a simple form of vector linear codes, called vector cyclic codes. It follows that the information theoretic optimum is bounded by the \emph{integrality gap} between the integer program and its LP relaxation. Section \ref{section:planar-graphs} shows that in the special case when the bipartite digraph is planar, the integrality gap is zero. In this case, optimality is achieved by a scalar cyclic code.  Section \ref{section:partial-clique-codes} considers a different representation of the original integer program that yields a smaller integrality gap. The dual problem of its LP relaxation leads to a more sophisticated \emph{partial clique coding strategy} that time-shares between Reed-Solomon erasure codes. The smaller integrality gap ensures that these codes are closer to the lower bound. These results provide new insight on the index coding problem and suggest that good codes can be found by exploring the LP relaxations of the tightest lower bound problem.
  
 \section{The weighted bipartite digraph} \label{section:weighted_bipartite_graph}
 
 There are $N$ receivers, also called \emph{users}. 
 Let $\mathcal{U} = \{u_1,\ldots,u_N\}$ be the set of users.   
 Assume there are $W$ total packets, labeled $\{q_1, \ldots, q_W\}$.  
 For each $m \in \{1, \ldots, W\}$, define $\mathcal{S}_m$ as the set of users in $\mathcal{U}$ that already have packet $q_m$ as side information, and  define $\mathcal{D}_m$ as the set of users in $\mathcal{U}$ that demand packet $q_m$.   
  Without loss of generality, assume that each packet is demanded by at least one user (else, that packet can be eliminated). 
  Thus, the demand set $\mathcal{D}_m$ is non-empty for all $m \in \{1, \ldots, W\}$.   On the other hand, the side information sets  $\mathcal{S}_m$ can be empty.  Indeed, the set $\mathcal{S}_m$ is empty if and only if no user has packet $q_m$ as side information. It is reasonable to assume that the set of users that demand a packet is disjoint from the set of users that already have that packet as side information, so that 
   $\mathcal{S}_m \cap \mathcal{D}_m = \emptyset$ for all $m \in \{1, \ldots, W\}$.

   This index coding problem is represented by a bipartite directed graph in \cite{NeelyTehraniZhang12}\cite{Tehrani12ISIT}, where user vertices are on the left of the graph, packet vertices are on the right, and the $\mathcal{S}_m$ and $\mathcal{D}_m$ sets are represented by directed arcs.   A directed graph is also called a \emph{digraph}. It is useful to extend this representation to a \emph{weighted} bipartite digraph as follows:  Two packets $q_k$ and $q_m$ are said to have the same \emph{type} if $\mathcal{S}_k = \mathcal{S}_m$ and $\mathcal{D}_k = \mathcal{D}_m$.  That is, two packets have the same type if they have the same side information and demand sets.  Note that if a user demands one packet of a certain type, then it must demand \emph{all} packets of that type.  Likewise, if a user has one packet of a certain type as side information, then it must have \emph{all} packets of that type. 
 
\begin{figure}[htbp]
   \centering
   \includegraphics[width=3.5in]{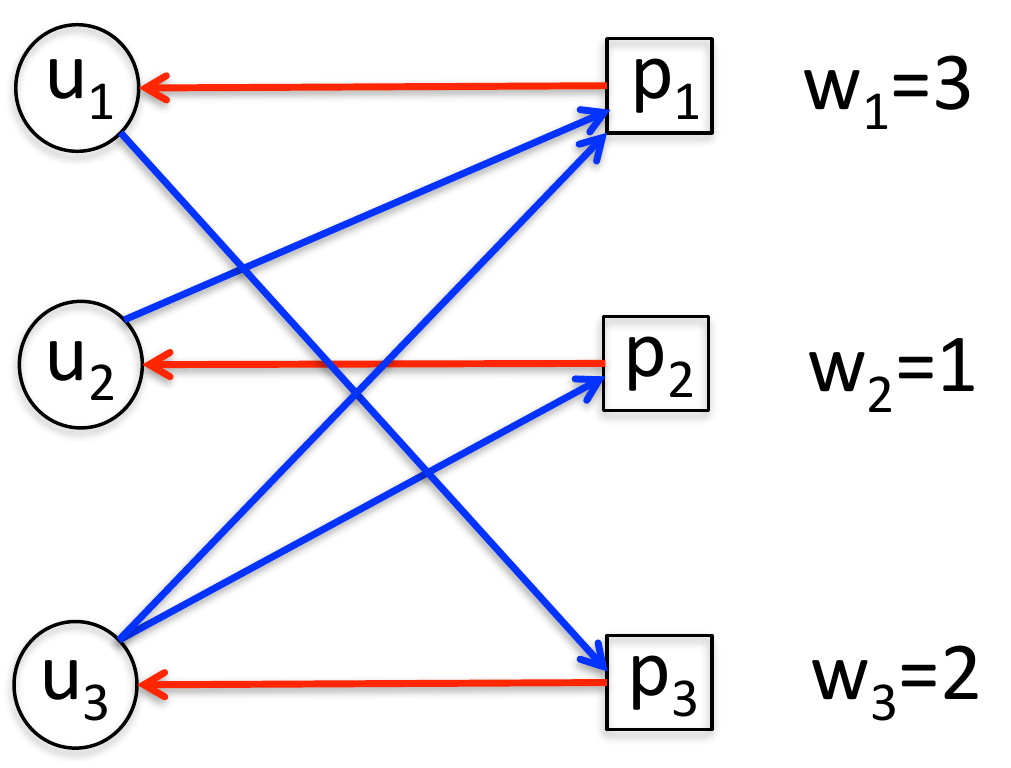} 
   \caption{The bipartite digraph representation of a unicast index coding problem with $3$ user vertices and $3$ packet type vertices. }
   \label{fig:bipartite_graph_example}
\end{figure}

   Let $M$ be the number of packet types, and let $\mathcal{P} = \{p_1, \ldots, p_M\}$ be the set of types.  
   The index coding problem can be represented by a weighted  bipartite digraph $\mathcal{G} =(\mathcal{U},\mathcal{P}, \mathcal{A},\mathcal{W}_{\mathcal{P}})$ as follows:  Let $\mathcal{U}$ be the set of vertices on the left side of the graph and let 
   $\mathcal{P}$ be the set of vertices on the right side of the graph (see Fig. \ref{fig:bipartite_graph_example}).  
   The arc set $\mathcal{A}$ has a user-to-packet arc $(u_n,p_m)$ if and only if user $u_n\in\mathcal{U}$ has all packets 
   of type $p_m$.  The arc set $\mathcal{A}$ has a packet-to-user arc $(p_m, u_n)$ if and only if user $u_n\in\mathcal{U}$ demands all packets of type $p_m$.  Finally, define   $\mathcal{W}_{\mathcal{P}}$ as the set of integral weights associated with packet vertices in $\mathcal{P}$.   The weight $w_{p_m} \in \mathcal{W}_{\mathcal{P}}$ of packet vertex $p_m \in \mathcal{P}$ 
   is equal to the number of packets of type $p_m$.   Thus, the total number of packets $W$ satisfies $W = \sum_{m=1}^M w_{p_m}$.

  A packet is said to be a \emph{unicast packet} if it is demanded by only one user, and is said to be a \emph{multicast packet} if it is demanded by two or more users.  An index coding problem is said to be \emph{unicast} if all packets are unicast packets. 
   The first index coding problem introduced by Birk and Kol in \cite{Birk98INFOCOM} was a unicast problem. The current paper also focuses exclusively on the unicast case.  
      Figure \ref{fig:bipartite_graph_example} shows an example of the weighted bipartite digraph representation for a unicast index coding problem with $3$ user vertices and $3$ packet types. In this example, packet types $p_1, p_2, p_3$ are demanded by users $u_1$, $u_2$, $u_3$, respectively, so that $\mathcal{D}_1 = \{u_1\}$, $\mathcal{D}_2 = \{u_2\}$, $\mathcal{D}_3 = \{u_3\}$.  Furthermore, the side information sets are as follows: 
      \begin{itemize} 
      \item Packet type $p_1$ is contained as side information by users in the set $\mathcal{S}_1 = \{u_2, u_3\}$. 
      \item Packet type $p_2$ is contained as side information by the user in the set $\mathcal{S}_2 = \{u_3\}$. 
      \item Packet type $p_3$ is contained as side information by the user in the set $\mathcal{S}_3 = \{u_1\}$. 
      \end{itemize}

The index coding problem with graph $\mathcal{G} =(\mathcal{U},\mathcal{P}, \mathcal{A},\mathcal{W}_{\mathcal{P}})$ can equally represent a system with $M$ variable size packets, where $w_{p_m}$ is the (integer) size of packet $p_m$. With this interpretation, each packet type represents a single packet.  Thus, this paper often refers to packet type $p_m$ as \emph{packet $p_m$}.  
   
\section{The acyclic subgraph bound and its LP relaxation} \label{section:integer_program_lower_bound}

The following definitions from graph theory are useful. A sequence of vertices $\{s_1,s_2,\ldots,s_K\}$ of a general digraph is defined as a \emph{cycle} if $(s_i,s_{i+1})\in \mathcal{A}$ for all $i\in\{1,2,\ldots,K-1\}$, all vertices in $\{s_1,s_2,\ldots,s_{K-1}\}$ are distinct,  and $s_1 = s_K$. A digraph is \emph{acyclic} if it contains no cycle.  A set of vertices is called a \emph{feedback vertex set} if the removal of vertices in this set leaves an acyclic digraph. 
In a vertex-weighted digraph, the feedback vertex set with the minimum sum weight is called the \emph{minimum feedback vertex set}. 

For the weighted bipartite digraph $\mathcal{G} =(\mathcal{U},\mathcal{P}, \mathcal{A},\mathcal{W}_{\mathcal{P}})$ (as defined in the previous section), there exists a subset $\mathcal{P}_{\text{fd}} \subseteq \mathcal{P}$ such that the removal of vertices in $\mathcal{P}_{\text{fd}}$ and all the associated packet-to-user arcs and user-to-packet arcs leaves an acyclic subgraph. In this case, $\mathcal{P}_{\text{fd}}$ is called a \emph{feedback packet vertex set}.  A trivial feedback packet vertex set is $\mathcal{P}_{\text{fd}} = \mathcal{P}$ and the corresponding acyclic subgraph has no packet vertex.   This trivial feedback packet vertex set has \emph{weight $W$}, since the sum weight of all packet vertices is $W$.  It is often possible to find a feedback packet vertex set with sum weight smaller than $W$.  The feedback packet vertex set with the minimum sum weight is called the \emph{minimum feedback packet vertex set}.  The acyclic subgraph induced by the deletion of the minimum feedback packet vertex set is called the \emph{maximum acyclic subgraph}. 

Assume that each transmission from the base station sends a number of bits equal to the number of bits in each of the fixed length packets.  It is trivial to satisfy all demands with $W$ transmissions, where each of the $W$ packets is successively transmitted without coding.   However, coding can often be used to reduce the number of transmissions.  Let $T_{\text{min}}(\mathcal{G})$ represent the minimum number of transmissions required to deliver all packets to their intended users for an index coding problem defined by the weighted bipartite digraph $\mathcal{G}$.  The value $T_{\text{min}}(\mathcal{G})$ considers all possible coding strategies.  A theorem in \cite{NeelyTehraniZhang12} provides an information theoretic lower bound on $T_{\text{min}}(\mathcal{G})$.  While the theorem holds for general (possibly multicast) index coding problems, this paper uses it in the unicast case. 

\begin{Thm}[Paraphrased from Theorem 1  and Lemma 1 in \cite{NeelyTehraniZhang12}] \label{Thm:lower-bound}
Consider an index coding problem $\mathcal{G} =(\mathcal{U},\mathcal{P}, \mathcal{A},\mathcal{W}_{\mathcal{P}})$. Let $\mathcal{P_{\text{fd}}} \subseteq \mathcal{P}$ be a feedback packet vertex set and let $\mathcal{G}^\prime$ be the acyclic subgraph induced by the deletion of $\mathcal{P}_{\text{fd}}$. If $\sum_{p_m\in\mathcal{G}^\prime} w_{p_m} = W^\prime$, then $T_{\text{min}}(\mathcal{G}) \geq W^\prime$. 
\end{Thm}

Suppose the largest cycle in digraph $\mathcal{G}$ involves $L$ packet vertices. Define the set of all cycles in $\mathcal{G}$ as $\mathcal{C} = \bigcup_{i=1}^L \mathcal{C}_i$,  where $\mathcal{C}_{i} , i=2,\ldots,L$ is the set of all cycles involving $i$ packet vertices. These cycles can possibly overlap, i.e., some of them can share common vertices.  The number of cycles can possibly be exponential in the number of vertices of the graph.   The problem of identifying the tightest lower bound provided by Theorem \ref{Thm:lower-bound} can be formulated as an integer linear programming (ILP) problem as below:
\begin{align*}
\text{(P1)}
\begin{aligned}
\underset{x_m, m=1,\ldots,M}{\text{max}} \quad &\sum_{m=1}^M x_m w_{p_m}\\
\text{s.t.} \quad \quad &\sum_{m=1}^M x_m \mathbf{1}_{\{p_m \in C_{i}\}} \leq i-1, \quad \forall C_i \in \mathcal{C}_i, i=2,\ldots, L\\
			&x_m \in \{0,1\},\quad m=1,\ldots, M
\end{aligned}
\end{align*}
where $x_m\in\{0,1\}, m=1,\ldots,M$ indicates if packet vertex $p_m$ remains in the acyclic subgraph, objective function $\sum_{m=1}^M x_m w_{p_m}$ is the sum weight of the acyclic subgraph, $\mathbf{1}_{\{p_m \in C_i\}} $ is the indicator function which equals one if and only if packet vertex $p_m$ participates in cycle $C_i \in \mathcal{C}_i$, and $\sum_{m=1}^M x_m \mathbf{1}_{\{p_m \in C_i \}} \leq i-1$ is the constraint that for each cycle $C_i \in \mathcal{C}_i$, at most  $i-1$ packet vertices remains in the acyclic subgraph. This problem finds  the maximum packet weighted acyclic subgraph formed by packet vertex deletion.

The integer constraints of the above problem can be convexified to form the following \emph{linear programming (LP) relaxation}: 
\begin{align*}
\text{(P1$^\prime$)}
\begin{aligned}
\underset{x_m, m=1,\ldots,M}{\text{max}} \quad &\sum_{m=1}^M x_m w_{p_m}\\
\text{s.t.} \quad \quad &\sum_{m=1}^M x_m \mathbf{1}_{\{p_m \in C_{i}\}} \leq i-1, \quad \forall C_i \in \mathcal{C}_i, i=2,\ldots, L\\
			& 0 \leq x_m  \leq 1,\quad m=1,\ldots, M
\end{aligned}
\end{align*}
The only difference between problem (P1) and its relaxation (P1$^\prime$) is that the constraints $x_m \in \{0, 1\}$ are changed to 
$0 \leq x_m \leq 1$. 

Define $val(\text{P1})$ as the optimal objective function value of the integer program (P1), being the size of the maximum acyclic subgraph.  Theorem \ref{Thm:lower-bound} implies that $val(\text{P1}) \leq T_{\text{min}}(\mathcal{G})$.  The optimal objective function value for the relaxation (P1$^\prime$) can be written as $val(\text{P1}^\prime) = val(\text{P1}) + gap(\text{P1}^\prime, \text{P1})$, where $gap(\text{P1}^\prime, \text{P1}) = val(\text{P1}^\prime) - val(\text{P1})$ is the \emph{integrality gap} between the LP relaxation  (P1$^\prime$) and the integer program (P1).  Since the relaxation  (P1$^\prime$)  has less restrictive constraints, the value of $gap(\text{P1}^\prime, \text{P1})$ is always non-negative.  The next section proves constructively that: 
\[ val(\text{P1}) \leq T_{\min}(\mathcal{G}) \leq val(\text{P1}) + gap(\text{P1}^\prime, \text{P1})\]
Thus, the difference between the minimum clearance time and the maximum acyclic subgraph bound is bounded by the integrality gap $gap(\text{P1}^\prime, \text{P1})$.  Furthermore, Section \ref{section:planar-graphs} shows that $gap(\text{P1}^\prime, \text{P1})=0$ in special cases when the digraph $\mathcal{G}$ is planar.

\section{Cyclic Codes and Linear Programming Duality}\label{section:cyclic-code-duality}

Inspired by the observation that the lower bound in Theorem \ref{Thm:lower-bound} is closely connected with cycles in graph $\mathcal{G}$, this section considers \emph{cyclic codes} that exploit  cycles in $\mathcal{G}$.  It is shown that the problem of finding the optimal cyclic code is the dual problem of the LP relaxation (P1$^\prime$). Thus, the performance gap between the optimal cyclic code and the optimal index code is ultimately bounded by the integrality gap $gap(\text{P1}^\prime, \text{P1})$. 

\subsection{Cyclic Codes} 
Suppose there exists a cycle in $\mathcal{G}$ that involves  $K$ users $\{u_1,u_2,\ldots, u_K\}$ and $K$ packets of the same size $\{q_1,q_2,\ldots,q_K\}$. In this cycle, user $u_1$ has $q_K$ as side information and demands $q_1$, user $u_2$ has $q_1$ as side information and demands $q_2$, user $u_3$ has $q_2$ as side information and demands $q_3$, and so on. If the weight of each packet node is identically one, a \textit{$K$-cycle coding action} can deliver all $K$ packets by transmitting $Z_i = q_i+q_{i+1}, i=1,\ldots,K-1$ with $K-1$ transmissions, where addition is the mod-$2$ summation of each bit in both packets. After transmissions, user $u_i\in\{u_2,\ldots,u_K\}$ can decode packet $q_{i}$ by performing $q_{i-1}+ Z_{i-1} = q_{i-1} + (q_{i-1}+q_{i}) = q_i$. At the same time, user $u_1$ can decode packet $q_1$ by performing: 
\begin{align*}
Z_1+\ldots+Z_{K-1} +q_K &=(q_1+q_2)+(q_2+q_3)+\ldots+(q_{K-1}+q_{K}) + q_K\\
					  &=q_1. 
\end{align*}

The linear index code of $\mathcal{G}$ is said to be cyclic if it uses a sequence of coding actions that involve 
only cyclic coding actions and direct broadcasts without coding. Linear codes can be further categorized into \emph{scalar linear codes} and \emph{vector linear codes} according to whether the transmitted message is a linear combination of the original packets or the subpackets obtained by subdivisions. In scalar linear codes, each packet is considered as an element of a finite field and the transmitted message is a linear combination of packets over that field. In vector linear codes, each packet is assumed to be sufficiently large and can be divided into many smaller subpackets and the transmitted message is a linear combination of  these subpackets instead of the original packets. The problem of finding the optimal scalar cyclic code to clear $\mathcal{G}$ can be formulated as an ILP problem as below:
\begin{align*}
\text{(P2)}
\begin{aligned}
\underset{\begin{subarray}{c}y_{C_i}, \forall C_i \in \mathcal{C}_i, i=2,\ldots,L; \\ y_m, m=1,\ldots,M\end{subarray}}{\text{min}} \quad &\sum_{i=2}^L \sum_{C_i \in \mathcal{C}_i} y_{C_i} (i-1) + \sum_{m=1}^M y_m\\
\text{s.t.} \quad \quad &y_m + \sum_{i=2}^L \sum_{C_i \in \mathcal{C}_i} y_{C_i} \mathbf{1}_{\{p_m \in C_i \}} \geq w_{p_m},\quad m=1,\ldots, M\\
			&y_{C_i} \text{ non-negative integral},\quad \forall C_i \in \mathcal{C}_i, i=2,\ldots, L\\
			&y_m \text{ non-negative integral}, \quad m=1,\ldots,M
\end{aligned}
\end{align*}
where $y_{C_i}$ is the number of cycle codes over each cycle $C_i,  \forall C_i \in \mathcal{C}_i, i=2,\ldots,L$, $y_m$ is the number of direct broadcasts over each packet vertex $p_m, m=1,\ldots,M$, objective function $\sum_{i=2}^L \sum_{C_i \in \mathcal{C}_i} y_{C_i} (i-1) + \sum_{m=1}^M y_m$ is the total number of transmissions, and $y_m + \sum_{i=2}^L \sum_{C_i \in \mathcal{C}_i} y_{C_i} \mathbf{1}_{\{p_m \in C_i\}} \geq w_{p_m}$ is the constraint that all the $w_{p_m}$ packets represented by packet vertex $p_m$ are cleared by either cyclic codes or direct broadcasts.

The LP relaxation of integer program (P2) is below:
\begin{align*}
\text{(P2$^\prime$)}
\begin{aligned}
\underset{\begin{subarray}{c}y_{C_i}, \forall C_i \in \mathcal{C}_i, i=2,\ldots,L; \\ y_m, m=1,\ldots,M\end{subarray}}{\text{min}} \quad &\sum_{i=2}^L \sum_{C_i \in \mathcal{C}_i} y_{C_i} (i-1) + \sum_{m=1}^M y_m\\
\text{s.t.} \quad \quad &y_m + \sum_{i=2}^L \sum_{C_i \in \mathcal{C}_i} y_{C_i} \mathbf{1}_{\{p_m \in C_i \}} \geq w_{p_m},\quad m=1,\ldots, M\\
			&y_{C_i} \geq 0 ,\quad \forall C_i \in \mathcal{C}_i, i=2,\ldots, L\\
			&y_m \geq 0, \quad m=1,\ldots,M
\end{aligned}
\end{align*}
The only difference between the above problem and the original problem (P2) is that the constraints that $y_{C_i}$ and $y_m$ are non-negative integral are replaced by the relaxed constraints that $y_{C_i} \geq 0$ and $y_m \geq 0$. 

Since all the parameters in the linear constraints of (P2$^\prime$) are integers, an optimal solution can be found that has all variables equal to rational numbers. Let an optimal solution of (P2$^\prime$) be $y^\ast_{C_i}, \forall C_i \in \mathcal{C}_i, i=2,\ldots,L;  y^\ast_m, m=1,\ldots,M$, and assume these values are all rational numbers.  The optimal vector cyclic code can be constructed as follows.  First, one can find an integer $\theta$ such that $\theta y^\ast_{C_i}, \forall C_i \in \mathcal{C}_i, i=2,\ldots,L;  \theta y^\ast_m, m=1,\ldots,M$ are all integers. Next, divide each packet into $\theta$ subpackets. After the subdivision, a single cyclic coding action over a cycle $C_i$  is no longer a linear combination of packets but a linear combination of subpackets.  Further, a single (uncoded) direct broadcast from a packet vertex $p_m$ is no longer the broadcast of one packet but one subpacket. Then, the optimal vector cyclic code performs $\theta y^\ast_{C_i}$ cyclic coding actions over each cycle $C_i, \forall C_i \in \mathcal{C}_i, i=2,\ldots,L$ and broadcasts $\theta y^\ast_m$ subpackets over each packet vertex $p_m, m=1,\ldots,M$.

Define $gap(\text{P2}, \text{P2}^\prime)$ as the integrality gap between integer program (P2) and its LP relaxation (P2$^\prime$). Since the relaxation (P2$^\prime$)  has less restrictive constraints, the value of $gap(\text{P2}, \text{P2}^\prime)$ is always non-negative. Let $T_{\text{cyclic}}(\mathcal{G})$ and $T^\prime_{\text{cyclic}}(\mathcal{G})$ be the clearance time attained by the optimal vector cyclic code and the optimal scalar cyclic code, respectively. Then $T_{\text{cyclic}}(\mathcal{G}) - T^\prime_{\text{cyclic}}(\mathcal{G}) = gap(\text{P2}, \text{P2}^\prime)$. 

\subsection{Duality Between Information Theoretical Lower Bounds and Cyclic Codes}

The duality between the tightest lower bound given by Theorem \ref{Thm:lower-bound} and the optimal cyclic code is formally stated in the following lemma. 

\begin{Lem}\label{Thm:LP-duality} The LP relaxations (P1$^\prime$) and (P2$^\prime$) form a primal-dual linear programming pair.  In particular, the vector cyclic code\footnote{Similarly, the scalar cyclic code associated with problem (P2) achieves a clearance time of $val(\text{P1}) + gap(\text{P1}^\prime, \text{P1}) + gap(\text{P2}, \text{P2}^\prime)$.} associated with problem (P2$^\prime$) achieves a clearance time of $val(\text{P1}) + gap(\text{P1}^\prime,\text{P1})$.
\end{Lem}

\begin{IEEEproof} 
The Lagrangian function of (P2$^\prime$) can be written as 
\begin{align*}
L(y_{C_i}, y_m, \lambda_m, \mu_{C_i}, \mu_m) &= \sum_{i=2}^L \sum_{C_i \in \mathcal{C}_i} y_{C_i} (i-1) + \sum_{m=1}^M y_m + \sum_{m=1}^M \lambda_m \big[ w_{p_m} - y_m - \sum_{i=2}^L \sum_{C_i\in \mathcal{C}_i} y_{C_i} \mathbf{1}_{\{p_m \in C_i\}} \big]  \\
&~~~- \sum_{i=2}^L\sum_{C_i\in \mathcal{C}_i} \mu_{C_i} y_{C_i} - \sum_{m=1}^M \mu_m y_m\\
&= \sum_{m=1}^M \lambda_m w_{p_m} + \sum_{i=2}^L \sum_{C_i\in\mathcal{C}_i} y_{C_i} \big[ (i-1) - \sum_{m=1}^M \lambda_m \mathbf{1}_{\{p_m \in C_i\}} - \mu_{C_i}\big] \\
&~~~+\sum_{m=1}^M y_m[1 - \lambda_m -\mu_m]
\end{align*}
where $\lambda_m \geq 0, m=1,\ldots,M$; $\mu_{C_i}\geq 0, \forall C_i \in \mathcal{C}_i, i=2,\ldots,L$ and $\mu_m \geq 0, m=1,\ldots,M$. The dual problem of (P2$^\prime$) is defined as:
\begin{align*}
\underset{\begin{subarray}{c}\lambda_m \geq 0, m=1,\ldots,M;\\\mu_{C_i}\geq 0, \forall C_i \in \mathcal{C}_i, i=2,\ldots,L;\\ \mu_m \geq 0, m=1,\ldots,M\end{subarray}}{\max}  \underset{\begin{subarray}{c} y_{C_i}\in \mathbb{R}, \forall C_i \in \mathcal{C}_i, i=2,\ldots,L\\ y_m \in \mathbb{R}, m=1,\ldots,M\end{subarray}} {\min}  L(y_{C_i}, y_m, \lambda_m, \mu_{C_i}, \mu_m) 
\end{align*}
Note that, 
\begin{align*}
\underset{\begin{subarray}{c} y_{C_i}\in \mathbb{R}, \forall C_i \in\mathcal{C}_i,i=2,\ldots,L\\ y_m \in \mathbb{R}, m=1,\ldots,M\end{subarray}} {\min}  L(y_{C_i}, y_m, \lambda_m, \mu_{C_i}, \mu_m)  = \left\{ \begin{array}{cl}
\sum_{m=1}^M \lambda_m w_{p_m} &  \text{if }\begin{subarray}{l} (i-1) - \sum_{m=1}^M \lambda_m \mathbf{1}_{\{p_m \in C_i\}} - \mu_{C_i}=0,\\ \forall C_i \in \mathcal{C}_i, i=2,\ldots,L\\ 1 - \lambda_m -\mu_m=0, m=1,\ldots,M \end{subarray}\\-\infty & \text{otherwise}\end{array}\right.
\end{align*}
Then, the dual problem of (P2$^\prime$) can be written as,
\begin{align*}
\begin{aligned}
\underset{\begin{subarray}{c}\lambda_m, m=1,\ldots,M;\\\mu_{C_i}, \forall C_i \in \mathcal{C}_i, i=2,\ldots,L;\\ \mu_m, m=1,\ldots,M\end{subarray}}{\text{max}} \quad &\sum_{m=1}^M \lambda_m w_{p_m}\\
\text{s.t.} \quad \quad &(i-1) - \sum_{m=1}^M \lambda_m \mathbf{1}_{\{p_m \in C_i\}} - \mu_{C_i}=0, \quad \forall C_i \in \mathcal{C}_i, i=2,\ldots,L\\
			&1 - \lambda_m -\mu_m=0, \quad m=1,\ldots,M\\
			&\lambda_m \geq 0, \quad m=1,\ldots,M\\
			&\mu_{C_i}\geq 0, \quad \forall C_i \in \mathcal{C}_i, i=2,\ldots,L\\
			&\mu_m \geq 0, \quad m=1,\ldots,M
\end{aligned}
\end{align*}
Eliminating variables $\mu_{C_i}, \forall C_i \in \mathcal{C}_i, i=2,\ldots,L$ and $\mu_m , m=1,\ldots,M$, we obtain
\begin{align*}
\begin{aligned}
\underset{\lambda_m, m=1,\ldots,M}{\text{max}} \quad &\sum_{m=1}^M \lambda_m w_{p_m}\\
\text{s.t.} \quad \quad &\sum_{m=1}^M \lambda_m \mathbf{1}_{\{p_m \in C_i\}} \leq (i-1), \quad \forall C_i \in \mathcal{C}_i, i=2,\ldots,L\\
			&0 \leq \lambda_m \leq 1, \quad m=1,\ldots,M\\
\end{aligned}
\end{align*}
The above problem is the same as (P1$^\prime$). Thus, the clearance time of the vector cyclic code associated with problem (P2$^\prime$) is equal to the value of the optimal objective function in problem (P1$^\prime$), which is $val(\text{P1}) + gap(\text{P1}^\prime, \text{P1})$.  Then, the clearance time of the scalar cyclic code associated with problem (P2) is equal to $val(\text{P1}) + gap(\text{P1}^\prime, \text{P1}) + gap(\text{P2}, \text{P2}^\prime)$.
\end{IEEEproof}

Thus far, we have proven the following lower and upper bound for the minimum clearance time of an index coding problem.  
\[ val(\text{P1}) \leq T_{\min}(\mathcal{G}) \leq val(\text{P1}) + gap(\text{P1}^\prime, \text{P1}) \]
where the first inequality follows from Theorem \ref{Thm:lower-bound} and the second inequality follows
from Lemma \ref{Thm:LP-duality}. 
Hence, the performance gap between the optimal index code and the optimal vector cyclic code is ultimately bounded by the integrality gap between integer program (P1) and its LP relaxation (P1$^\prime$).

There are various techniques for bounding the integrality gaps of integer linear programs, such as the random rounding methods in \cite{Raghavan87Combinatorica,Raghavan88}.  Rather than explore this direction, the next section provides a special case where the gap is equal to zero.

\section{Optimality of Cyclic Codes in Planar Bipartite Graphs} \label{section:planar-graphs} 

In graph theory, a \emph{planar graph} is a graph that can be drawn as a picture on a 2-dimensional plane in a way so  that no two arcs meet at a point other than a common vertex.  The main result in this section is the following theorem: 

\begin{Thm} \label{Thm:cyclic_code_planar_graph}
If the bipartite digraph $\mathcal{G}$ for a (unicast) index coding problem is planar, then $val(\text{P1}) = val(\text{P2})$, i.e., $gap(\text{P1}^\prime, \text{P1}) = 0$ and $gap(\text{P2}, \text{P2}^\prime) = 0$. Hence, the (scalar) cyclic code given by (P2) is an optimal index code. 
\end{Thm} 

The proof of Theorem \ref{Thm:cyclic_code_planar_graph} relies on the cycle-packing and feedback arc set duality in arc-weighted planar graphs, which is summarized in the following theorem.

\begin{Thm}[Paraphrased from Theorem 2.1 in \cite{Guenin11Combinatorica}  and originally proven in \cite{Younger78Minimax}]  \label{Thm:max-min-duality-planar}
Let $\mathcal{G}=(\mathcal{V},\mathcal{A}, \mathcal{W}_{\mathcal{A}})$ be an arc-weighed planar digraph where $\mathcal{V}$ is the set of vertices, $\mathcal{A}$ is the set of arcs and $\mathcal{W}_{\mathcal{A}}$ is an integral arc weight assignment which assigns each arc $a \in \mathcal{A}$ a non-negative integral weight $w_a \in \mathbb{Z}^+$. Let $\mathcal{C}$ be the set of cycles in $\mathcal{G}$.  We have
\begin{align} 
&\min\Big\{\sum_{a\in \mathcal{A}}  x_a w_a: \sum_{a \in \mathcal{A}} x_a \mathbf{1}_{\{a\in C\}}\geq 1, \forall C\in \mathcal{C}; x_{a}\in\{0,1\},\forall a \in \mathcal{A}\Big\} \nonumber \\
=&\max\Big\{\sum_{C\in\mathcal{C}} y_C: \sum_{C\in\mathcal{C}} y_C \mathbf{1}_{\{a\in C\}} \leq w_a, \forall a\in \mathcal{A}; y_C\in \mathbb{Z}^+, \forall C\in \mathcal{C} \Big\}. \label{eq:Guenin} 
\end{align}
\end{Thm}

The integer program on the left-hand-side of \eqref{eq:Guenin} is a minimum feedback arc set problem, while the integer program on the right-hand-side of \eqref{eq:Guenin} is a cycle packing problem. Both problems are associated with \emph{arc weighted digraphs}. To apply this theorem, we introduce the respective complementary problems of (P1) and (P2). The complementary problem of (P1) is a minimum feedback \emph{packet vertex} set problem and the complementary problem (P2) is a cycle packing problem. However, both complementary problems are associated with \emph{packet-vertex-weighted digraphs}. To settle this issue,  we modify the bipartite digraph $\mathcal{G}$ to produce an arc-weighted digraph $\mathcal{G}^s$, which is planar if and only if $\mathcal{G}$ is planar.  We then show that the minimum feedback packet vertex set problem and the cycle packing problem in $\mathcal{G}$ can be reduced to the minimum feedback arc set problem and the cycle packing problem in $\mathcal{G}^s$, respectively.  The following subsections develop the proof of Theorem \ref{Thm:cyclic_code_planar_graph} and provide some additional consequences. 

\subsection{Complementary Problems}

The integer program (P1) finds the maximum packet weighted acyclic subgraph.    This is equivalent to finding the minimum weight feedback packet vertex set.  Indeed, this is the set of packets whose deletion induce the maximum packet weighted acyclic subgraph. Thus, an equivalent problem to (P1) is: 

\begin{align*}
\text{(P3)}
\begin{aligned}
\underset{x_m, m=1,\ldots,M}{\text{min}} \quad &\sum_{m=1}^M x_m w_{p_m}\\
\text{s.t.} \quad \quad &\sum_{m=1}^M x_m \mathbf{1}_{\{p_m \in C_i\}} \geq 1,  \quad \forall C_i \in \mathcal{C}_i, i=2,\ldots, L\\
			&x_m \in \{0,1\},\quad m=1,\ldots, M
\end{aligned}
\end{align*}
where $x_m\in\{0,1\}, m=1,\ldots,M$ indicates if packet vertex $p_m$ is selected into the feedback vertex set, objective function $\sum_{m=1}^M x_m w_{p_m}$ is the sum weight of the feedback vertex set, $\mathbf{1}_{\{p_m \in \mathcal{C}_i\}} $ is the indicator function which equals one only if packet vertex $p_m$ participates in cycle $C_i, \forall C_i \in \mathcal{C}_i, i=2,\ldots,L$, and $\sum_{m=1}^M x_m \mathbf{1}_{\{p_m \in C_i\}} \geq 1$ is the constraint that at least one packet vertex in each cycle is selected into the feedback vertex set. If $x^\ast_m, m=1,\ldots,M$ is the optimal solution of (P3) and attains the optimal value $W_0$, then $\overline{x}^\ast_m = 1- x^\ast_m, m=1,\ldots,M$ is the optimal solution of (P1) and attains the optimal value $ W- W_0$.

Now consider the integer program related to cyclic coding.  
It is now useful to write the complementary problem to the cyclic coding problem (P2).  
 In \cite{Chaudhry11}, Chaudhry et. al. introduced the concept of complementary index coding problems. Instead of trying to find the minimum number of transmissions to clear the problem, the complementary index coding problem is formulated to maximize the number of saved transmissions by exploiting a specific code structure. 
Recall that  any $K$-cycle code can deliver $K$ packets in $K-1$ transmissions and hence one transmission is saved in each $K$-cycle code. If the weight of each packet is not identically one, then $K$-cycle coding actions can be performed $w_{\min} = \min\{w_{p_1},\ldots,w_{p_K}\}$ times on the same cycle. By performing $K$-cycle coding $w_{min}$ times and then directly broadcasting the remaining packets (uncoded), the base station can deliver $w_{\text{total}} = \sum_{k=1}^K w_{p_k}$ packets with $w_{\text{total}} -w_{\min}$ transmissions.Thus, i$w_{\min}$ transmissions are saved. 

The complementary index coding problem which aims to maximize the number of saved transmissions by exploiting scalar cycles in $\mathcal{G}$ can be formulated as an ILP problem as below:

\begin{align*}
\text{(P4)}
\begin{aligned}
\underset{y_{C_i}, \forall C_i \in \mathcal{C}_i, i=2,\ldots,L}{\text{max}} \quad &\sum_{i=2}^L \sum_{C_i\in \mathcal{C}_i} y_{C_i}\\
\text{s.t.} \quad \quad &\sum_{i=2}^L \sum_{C_i\in \mathcal{C}_i} y_{C_i} \mathbf{1}_{\{p_m \in C_i\}} \leq w_{p_m},\quad m=1,\ldots, M\\
			&y_{C_i} \text{ non-negative integral},\quad \forall C_i \in \mathcal{C}_i, i=2,\ldots, L
\end{aligned}
\end{align*}
where $y_{C_i}$ is the number of cycle codes over each cycle $C_i \in \mathcal{C}_i, \forall C_i \in \mathcal{C}_i, i=2,\ldots,L$, objective function $\sum_{i=2}^L \sum_{C_i\in \mathcal{C}_i} y_{C_i}$ is the total number of cycle codes, i.e., total number of saved transmissions, and $\sum_{i=2}^L \sum_{C_i\in \mathcal{C}_i} y_{C_i} \mathbf{1}_{\{p_m \in C_i\}} \leq w_{p_m}$ is the constraint that each packet vertex $p_m$ can participate in at most  $w_{p_m}$ cycle codes. This is important because if packet $p_m$ has already participated $w_{p_m}$ times in cyclic coding actions, then all of its packets have been delivered and new cyclic coding actions that involve this packet vertex can no longer save any transmissions.  $K-1$ transmissions in this new cycle code clear $K-1$ packets for other packet vertices and $1$ useless duplicate packet for packet vertex $p_m$. No transmission is saved on this new cycle. If the optimal solution of (P4) is $y^\ast_{C_i},\forall C_i \in \mathcal{C}_i, i=2,\ldots,L$ and attains the optimal value $W_0$, then the optimal solution of (P2) is $\overline{y}^\ast_{C_i} = y^\ast_{C_i},\forall C_i \in \mathcal{C}_i, i=2,\ldots,L, \overline{y}^\ast_m = w_{p_m} - \sum_{i=2}^L \sum_{C_i\in \mathcal{C}_i} y^\ast_{C_i} \mathbf{1}_{\{p_m \in C_i\}} , m=1,\ldots,M$ and attains the optimal value $W - W_0$.

\subsection{Packet Split Digraphs}
\begin{Def}[Packet Split Digraphs] Given a graph $\mathcal{G} =(\mathcal{U},\mathcal{P}, \mathcal{A},\mathcal{W}_{\mathcal{P}})$, we construct the corresponding packet split digraph $\mathcal{G}^{\text{s}} = (\mathcal{V}^{\text{s}}, \mathcal{A}^{\text{s}}, \mathcal{W}^{\text{s}})$ as follows:
\begin{enumerate}
\item For each packet vertex $p_m \in \mathcal{P}, m=1,\ldots, M$, we create two packet vertices $p_m^{\text{in}}$ and $p_m^{\text{out}}$. Let $\mathcal{V}^{\text{s}} = \mathcal{U} \cup \{p_1^{\text{in}},p_1^{\text{out}},p_2^{\text{in}},p_2^{\text{out}},\ldots,p_M^{\text{in}},p_M^{\text{out}}\}$.
\item For each packet vertex  $p_m \in \mathcal{P}, m=1,\ldots, M$, we create a packet-to-packet arc $(p_m^{\text{in}},p_m^{\text{out}})$ in $\mathcal{A}^{\text{s}}$. For each arc $(u_n,p_m)\in \mathcal{A}$, we create a user-to-packet arc $(u_n, p_m^{\text{in}})$ in  $\mathcal{A}^{\text{s}}$. For each arc $(p_m,u_n) \in \mathcal{A}$, we create a packet-to-user arc $(p_m^{\text{out}},u_n)$ in $\mathcal{A}^{\text{s}}$.
\item For each arc $(p_m^{\text{in}},p_m^{\text{out}})$ in $\mathcal{A}^{\text{s}}$, we assign a weight which is equal to $w_{p_m} \in \mathcal{W}_{\mathcal{P}}$. For each arc $(u_n, p_m^{\text{in}})$ or $(p_m^{\text{out}},u_n)$ in $\mathcal{A}^{\text{s}}$, we assign an integral weight which is larger than $\sum_{m=1}^M w_{p_m}$.
\end{enumerate}
\end{Def}

\begin{figure}
   \centering
   \includegraphics[width=3in]{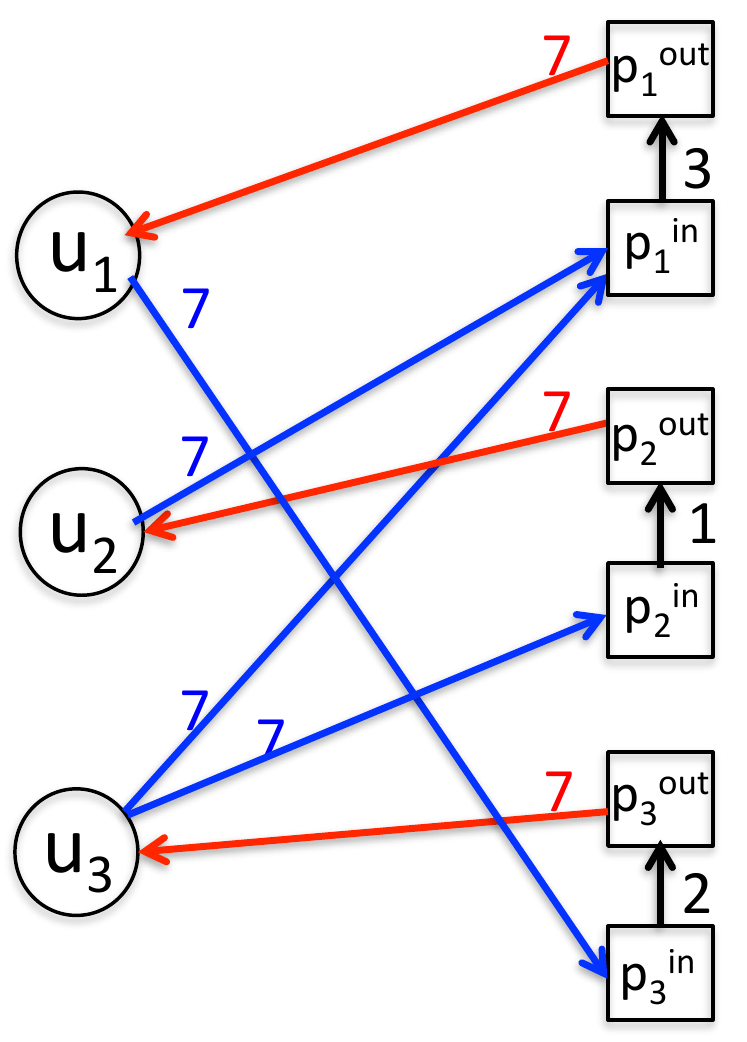} 
   \caption{The packet split digraph constructed from the bipartite digraph given in Figure \ref{fig:bipartite_graph_example} }
   \label{fig:packet_split_graph}
\end{figure}

For any bipartite digraph $\mathcal{G}$, the packet split digraph $\mathcal{G}^s$, which is an arc-weighted digraph, can always be constructed. Figure \ref{fig:packet_split_graph} shows the packet split digraph constructed from the bipartite digraph in Figure \ref{fig:bipartite_graph_example}. In any digraph, a set of arcs is called a feedback arc set if the removal of arcs in this set leaves an acyclic digraph. If the digraph is arc-weighted, the feedback arc set with the minimum sum weight is called the minimum feedback arc set.

The following facts summarize the connections between the packet split digraph and the original digraph.

\begin{Fact}
There is a bijection between $\mathcal{G}$ and $\mathcal{G}^{\text{s}}$. This bijection maps user vertices, user-to-packet arcs, packet vertices, and packet-to-user arcs in $\mathcal{G}$ to user vertices, user-to-packet arcs, packet-to-packet arcs, and packet-to-user arcs in $\mathcal{G}^s$, respectively. Thus, this bijection also maps cycles in $\mathcal{G}$ to cycles in $\mathcal{G}^{\text{s}}$.
\end{Fact}
\begin{IEEEproof}
The bijection can be easily identified according to the construction rule of the packet split digraph.
\end{IEEEproof}

\begin{Fact}
Every minimum feedback arc set of packet split graph $\mathcal{G}^s$ contains only packet-to-packet arcs and no packet-to-user arcs or user-to-packet arcs. 
\end{Fact}
\begin{IEEEproof}
In digraph $\mathcal{G}$, each cycle contains at least one packet vertex. By Fact 1, each cycle $\mathcal{G}^s$ contains at least one packet-to-packet arc. As such, the arc set composed of all packet-to-packet arcs is a feedback arc set of $\mathcal{G}^s$ and this feedback arc set contains no packet-to-user arcs or user-to-packet arcs. Note that the sum weight of this arc set is strictly less than the weight of any single packet-to-user or user-to-packet arc. Any feedback arc set with a packet-to-user arc or user-to-packet arc has a sum weight strictly larger than that of this one and hence can not be a minimum feedback arc set. 
\end{IEEEproof}
\begin{Fact}
If  $\mathcal{A}^{\text{s}}_{\text{fd}} \subseteq \mathcal{A}^{\text{s}}$ is a minimum feedback arc set of the packet split digraph $\mathcal{G}^{\text{s}}$, then a minimum feedback packet vertex set  $\mathcal{P}_{\text{fd}} \subseteq \mathcal{P}$ of $\mathcal{G}$ is immediate.  In addition, the sum weight of $\mathcal{P}_{\text{fd}}$ is equal to the sum weight of $\mathcal{A}^{\text{s}}_{\text{fd}}$.
\end{Fact}
\begin{IEEEproof}
Let $\mathcal{A}^\text{s}_{\text{fd}}$ be a minimum feedback arc set of $\mathcal{G}^s$ and the sum weight of $\mathcal{A}^\text{s}_{\text{fd}}$ be $W_{\text{fd}}$. By Fact 2, $\mathcal{A}^\text{s}_{\text{fd}}$ contains only packet-to-packet arcs. By Fact 1, the packet vertex set $\mathcal{P}_{\text{fd}} \subseteq \mathcal{P}$ composed by packet vertices corresponding to arcs in $\mathcal{A}^\text{s}_{\text{fd}}$ is a feedback packet vertex set of $\mathcal{G}$ and the sum weight of $\mathcal{P}_{\text{fd}}$ is equal to $W_{\text{fd}}$. If $\mathcal{P}_{\text{fd}}$ is not a minimum feedback packet vertex set, there must exist a minimum feedback packet vertex set, say $\mathcal{P}_{\text{fd}}^\prime$, whose sum weight $W_{\text{fd}}^\prime < W_{\text{fd}}$. By Fact 1, the counterpart of $\mathcal{P}_{\text{fd}}^\prime$ in $\mathcal{G}^s$ is a feedback arc set and the sum weight of this feedback arc set is equal to $W_{\text{fd}}^\prime$. Denote this feedback arc set as $\mathcal{A}_{\text{fd}}^{s,\prime}$, then $\mathcal{A}_{\text{fd}}^{s,\prime}$ has a sum weight strictly less than $W_{\text{fd}}$. This contradicts the fact that $\mathcal{A}^\text{s}_{\text{fd}}$ is a minimum feedback arc set of $\mathcal{G}^s$. Hence, $\mathcal{P}_{\text{fd}}$ must be a minimum feedback packet set of $\mathcal{G}$.
\end{IEEEproof}

\subsection{Optimality of Cyclic Codes in Planar Graphs}

\begin{figure}
   \centering
   \includegraphics[width=4.5in]{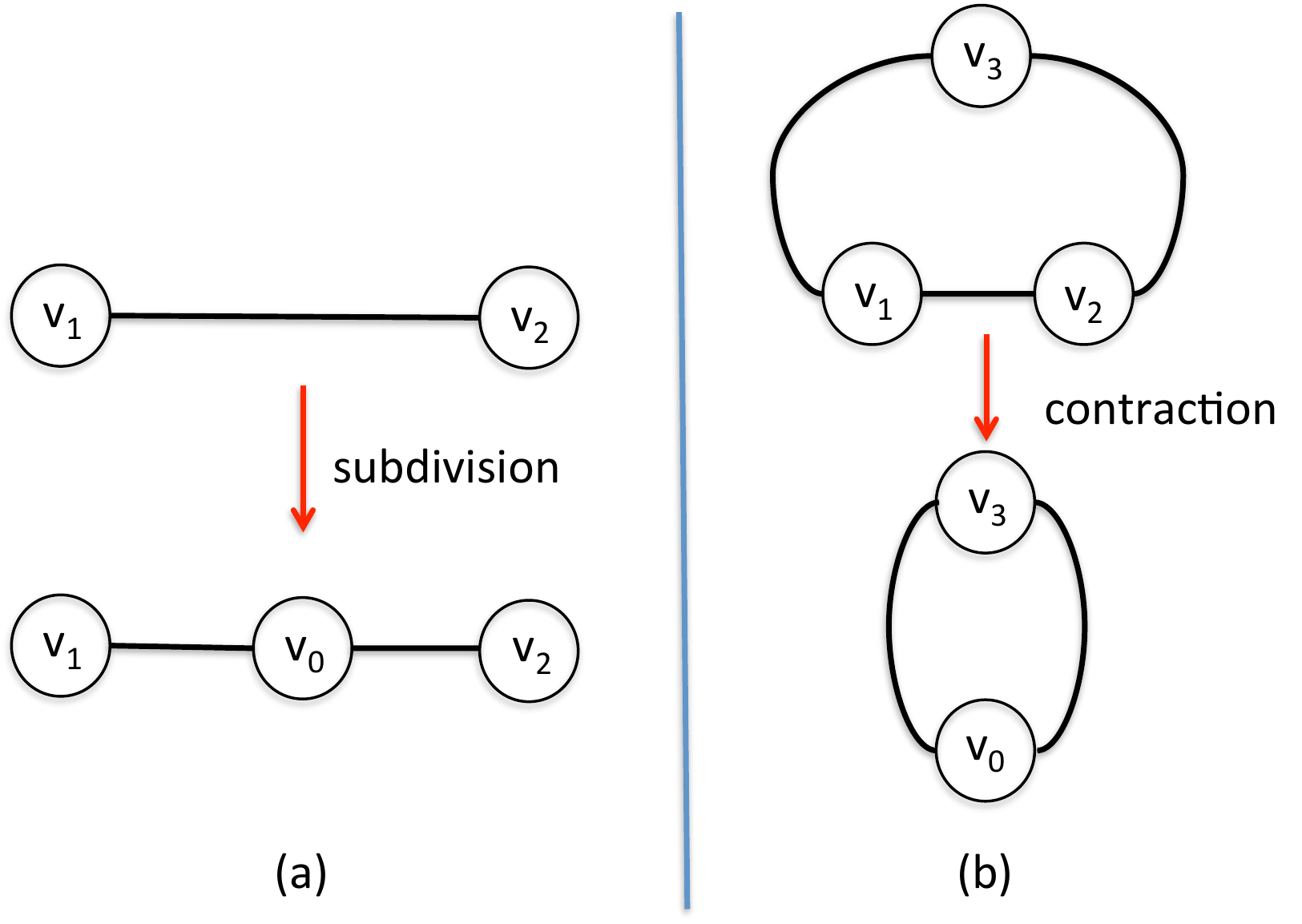} 
   \caption{(a) Subdivision of arc $(v_1,v_2)$. (b) Contraction of arc $(v_1,v_2).$ }
   \label{fig:subdivision_contraction}
\end{figure}

\begin{figure}
   \centering
   \includegraphics[width=4.5in]{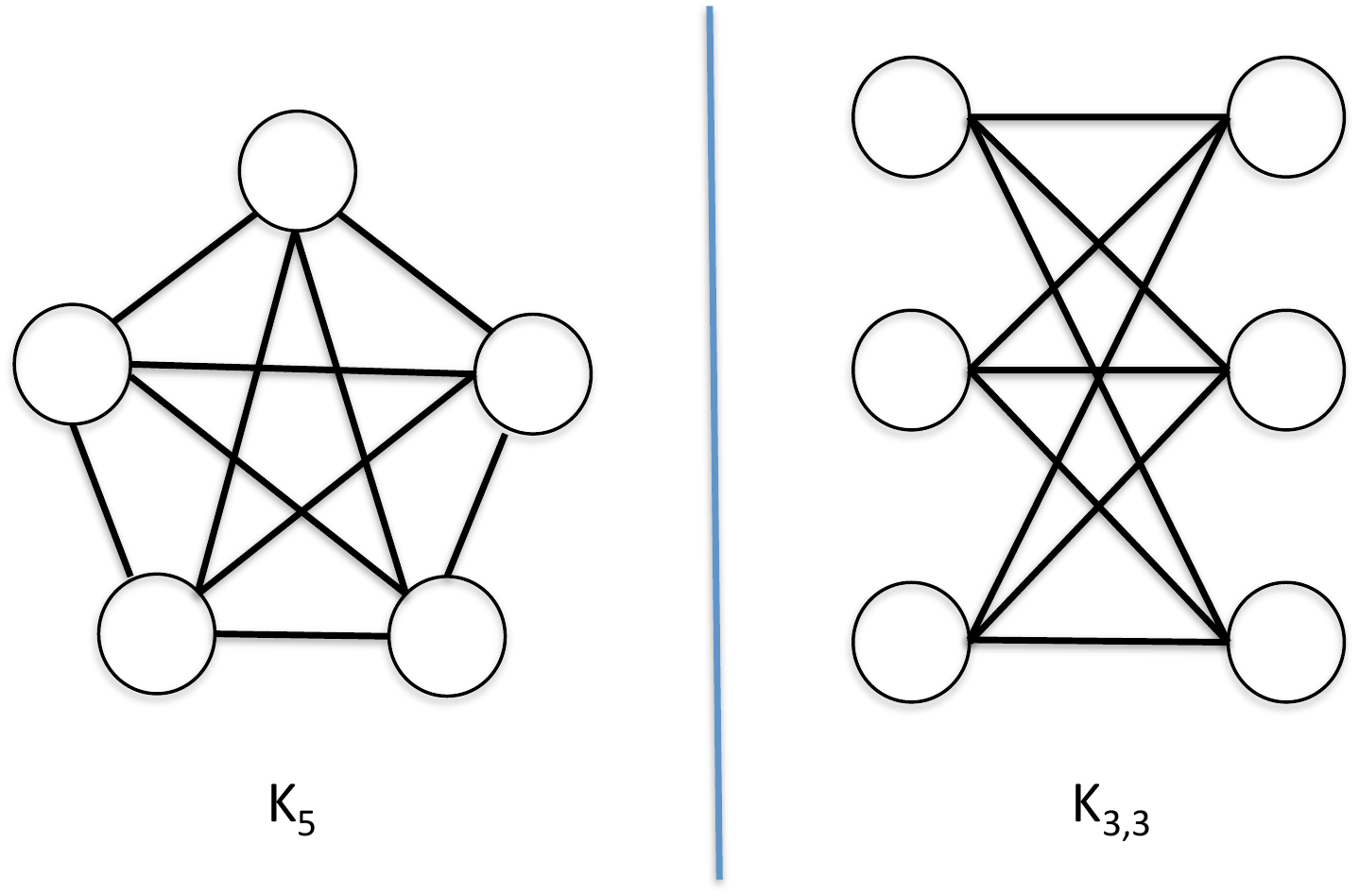} 
   \caption{$K_5$ and $K_{3,3}$ }
   \label{fig:k5k33}
\end{figure}

The planarity of a digraph is not affected by arc directions, so that a digraph is planar if and only if its undirected counterpart, where all directed arcs are turned into undirected edges,  is planar.   In an undirected graph, subdividing an edge  $(v_1,v_2)$ is the operation of deleting edge $(v_1,v_2)$, adding a vertex $v_0$, and adding edges $(v_1,v_0)$ and $(v_0,v_2)$ (see Figure \ref{fig:subdivision_contraction}a); contracting/shrinking an edge $(v_1,v_2)$ is the operation of deleting edge$(v_1,v_2)$, adding a vertex $v_0$, replacing any edge $(v,v_1)$ with $(v,v_0)$, and replacing any edge $(v_2,v)$ with $(v_0,v)$ (see Figure \ref{fig:subdivision_contraction}b). If a graph $\mathcal{G}$ is planar, subdividing and contracting operations preserve the planarity.  A graph $\mathcal{G}^\prime$ is said to be a subdivision of $\mathcal{G}$ if $\mathcal{G}^\prime$ is obtained from $\mathcal{G}$ by a sequence of edge subdividing operations. A graph $\mathcal{G}^\prime$ is said to be a minor of $\mathcal{G}$ if $\mathcal{G}^\prime$ is a subgraph of the graph obtained from $\mathcal{G}$ by a sequence of edge contracting operations. The simplest two non-planar graphs are the complete graph with $5$ vertices, which is denoted as $K_{5}$, and the complete bipartite graph with $3$ vertices on one side and $3$ vertices on the other side, which is denoted as $K_{3,3}$. Both of them are drawn in Figure \ref{fig:k5k33}.

The following theorem provides a sufficient and necessary condition for the planarity of an undirected graph.
\begin{Thm}[Page 24 in \cite{book_ModernGraphTheory} and originally proven by Wagner in 1937] \label{Thm:Wagner1937}
An undirected graph $\mathcal{G}$ is planar if and only if $\mathcal{G}$ contains neither $K_{5}$ nor $K_{3,3}$ as a minor.
\end{Thm}

In the index coding problem, a packet is said to be a \emph{uniprior packet} if it is contained as side information by only one user. The following lemma is proposed to characterize the planarity of the packet split graph $\mathcal{G}^s$.
\begin{Lem} \label{Lem:unicast-or-uniprior-planarity}
Let $\mathcal{G}$  be an index coding problem where each packet vertex is either unicast or uniprior and let $\mathcal{G}^s$ be the packet split digraph of $\mathcal{G}$.  $\mathcal{G}^{\text{s}}$ is planar if and only if $\mathcal{G}$ is planar.
\end{Lem} 
\begin{IEEEproof}
\begin{itemize}
\item
``only if" part: This part is relatively easy.  Assume $\mathcal{G}^{\text{s}}$ is planar and is drawn in a plane. A planar drawing of $\mathcal{G}$ can be obtained by contracting all the packet-to-packet arcs of $\mathcal{G}^{\text{s}}$ into packet vertices. This part holds for any $\mathcal{G}$ even if some packet vertex is neither unicast nor uniprior.
\item 
``if" part: Assume $\mathcal{G}$ is planar and is drawn  in a plane.  A planar drawing of $\mathcal{G}^{\text{s}}$ can be obtained by subdividing all packet-to-user arcs and user-to-packet arcs in $\mathcal{G}$. Specifically,  for each unicast packet vertex $p_m$ with one single outgoing link, we can subdivide the outgoing link into two parts; add a new node $p_m^{\text{out}}$ in the middle and reindex the node $p_m$ as $p_m^{\text{in}}$. Similarly, for each uniprior packet vertex $p_m$ with one single incoming link, we can subdivide the incoming link into two parts; add a new node $p_m^{\text{in}}$ in the middle and reindex the node $p_m$ as $p_m^{\text{out}}$. The subdivision operations as above yield a planar drawing of $\mathcal{G}^{\text{s}}$.
\end{itemize}  
\end{IEEEproof}

\begin{Cor}\label{Cor:unicast_planarity}
For any unicast index coding problem $\mathcal{G}$, $\mathcal{G}^s$ is planar if and only if $\mathcal{G}$ is planar.
\end{Cor}

Now we are ready to present the main result in this section.


{\it Theorem 2:} (Restated)
If the bipartite digraph $\mathcal{G}$ for a (unicast) index coding problem is planar, then $val(\text{P1}) = val(\text{P2})$, i.e., $gap(\text{P1}^\prime, \text{P1}) = 0$ and $gap(\text{P2}, \text{P2}^\prime) = 0$. Hence, the  cyclic code given by (P2) is an optimal index code.

\begin{IEEEproof}
Since $\mathcal{G}$ is a planar graph and this is a unicast index coding problem, $\mathcal{G}^s$ is also planar graph by Corollary \ref{Cor:unicast_planarity}. Let $\mathcal{G}^s = (\mathcal{V}^{\text{s}}, \mathcal{A}^{\text{s}}, \mathcal{W}^{\text{s}})$ be the packet spit digraph of $\mathcal{G} =(\mathcal{U},\mathcal{P}, \mathcal{A},\mathcal{W}_{\mathcal{P}})$. Let $\mathcal{C}^s$ be the set of cycles in $\mathcal{G}^s$.  The minimum feedback arc set problem in $\mathcal{G}^s$ can be formulated as an integer linear programming problem as follows:
\begin{align*}
\text{(P3$^\ast$)}
\begin{aligned}
\underset{x_a, a\in \mathcal{A}}{\text{min}} \quad &\sum_{a\in\mathcal{A}}^M x_a w_{a}\\
\text{s.t.} \quad \quad &\sum_{a\in\mathcal{A}} x_a \mathbf{1}_{\{a \in C\}} \geq 1, \quad \forall C\in \mathcal{C}^s\\
			&x_a \in \{0,1\},\quad a \in\mathcal{A}
\end{aligned}
\end{align*}
Similarly, the cycle-packing problem in $\mathcal{G}^s$ can formulated as another integer linear programming as follows:
\begin{align*}
\text{(P4$^\ast$)}
\begin{aligned}
\underset{y_{C}, C\in\mathcal{C}^s}{\text{max}} \quad &\sum_{C\in\mathcal{C}^s} y_C\\
\text{s.t.} \quad \quad &\sum_{C\in \mathcal{C}^s}^L y_C \mathbf{1}_{\{a \in C\}} \leq w_a,\quad \forall a \in \mathcal{A}^s\\
			&y_C \text{ non-negative integral},\forall C\in \mathcal{C}^s
\end{aligned}
\end{align*}

By Theorem \ref{Thm:max-min-duality-planar}, if $\mathcal{G}^s$ is a planar graph, then (P3$^\ast$) and (P4$^\ast$) have the same optimal value. In what follows, we show that the optimal value of (P3) is equal to that of (P3$^\ast$) and the optimal value of (P4) is equal to that of (P4$^\ast$). 
\begin{itemize}
\item {\bf (P3) and (P3$^\ast$) have the same optimal value:} By Fact 3, the minimum feedback arc set corresponding to the solution of (P3$^\ast$) can be converted to a minimum feedback packet set solution of (P3) which attains the same optimal objective function value as that of (P3$^\ast$). On the other hand, by Fact 1, the optimal solution of (P3) can be converted to a solution of (P3$^\ast$) which attains the same objective value as that of (P3).
\item {\bf (P4) and (P4$^\ast$) have the same optimal value:} By Fact 1, there is a bijection from $\mathcal{C}$ to $\mathcal{C}^s$. This is equivalent to say, there is a bijection from variables in (P4) to those in (P4$^\ast$). Let $\mathcal{A}^s_1$ be the set of packet-to-packet arcs and $\mathcal{A}^s_2$ be the set of packet-to-user and user-to-packet arcs. So $\mathcal{A}^s_1 \cup \mathcal{A}^s_2 = \mathcal{A}^s$ and $\mathcal{A}^s_1 \cap \mathcal{A}^s_2 = \emptyset$. The constraints $\sum_{C\in \mathcal{C}^s} y_C \mathbf{1}_{\{a \in C\}} \leq w_a, \forall a\in \mathcal{A}^s_1$ in (P4$^\ast$) are essentially the same as the constraints $\sum_{i=2}^L \sum_{C_i \in  \mathcal{C}_i} y_{C_i} \mathbf{1}_{\{p_m \in C_i\}} \leq w_{p_m},m=1,\ldots, M$ in (P4).  The other inequality constraints $\sum_{C\in \mathcal{C}^s} y_C \mathbf{1}_{\{a \in C\}} \leq w_a$ over $a \in \mathcal{A}^s_2$ can be shown to be redundant as follows.  Let $y_{C}, C\in \mathcal{C}^s$ be an arbitrary non-negative integral vector which satisfies all the constraints $\sum_{C\in \mathcal{C}^s} y_C \mathbf{1}_{\{a \in C\}} \leq w_a$ over $a \in \mathcal{A}^s_1$. Due to the bipartite property, each cycle in $\mathcal{G}$ contains at least one packet vertex. By Fact 1, each cycle in $\mathcal{G}^s$ contains at least one packet-to-packet arc. Thus, for any $C\in \mathcal{C}^s$, there exists some $a\in\mathcal{A}^s_1$ such that $\mathbf{1}_{\{a\in \mathcal{C}\}}  =1$. Then, for any $\bar{a}\in \mathcal{A}^s_2$ we have,
\begin{align*}
\sum_{C\in \mathcal{C}^s} y_C \mathbf{1}_{\{\bar{a} \in C\}} &\leq \sum_{C\in \mathcal{C}^s} y_C\\
											&\leq \sum_{C\in\mathcal{C}^s} \big[ y_C \cdot \sum_{a\in \mathcal{A}^s_1} \mathbf{1}_{\{a\in \mathcal{C}\}}\big] \\
											&=  \sum_{a\in \mathcal{A}^s_1} \big[ \sum_{C\in\mathcal{C}^s}  y_C \mathbf{1}_{\{a\in \mathcal{C}\}}\big]\\
											&\leq  \sum_{a\in \mathcal{A}^s_1} w_a\\
											&< w_{\bar{a}}
\end{align*}
where the first inequality follows from the fact that $0\leq \mathbf{1}_{\{\bar{a} \in C\}} \leq 1$; the second inequality follows from the fact that  for any $C\in \mathcal{C}^s$ there exists some $a\in\mathcal{A}^s_1$ such that $\mathbf{1}_{\{a\in \mathcal{C}\}}  =1$; the third inequality follows from the fact that all the constraints $\sum_{C\in \mathcal{C}^s} y_C \mathbf{1}_{\{a \in C\}} \leq w_a$ over $a \in \mathcal{A}^s_1$ are satisfied; and the last inequality follows from the fact that the weight of any packet-to-user arc or user-to-packet-arc is strictly larger than the sum weight of all packet-to-packet arcs. This is to say all the constraints $\sum_{C\in \mathcal{C}^s} y_C \mathbf{1}_{\{a \in C\}} \leq w_a$ over $a \in \mathcal{A}^s_1$ are automatically satisfied and hence redundant. Hence, (P4) and (P4$^\ast$) are two equivalent optimization problems.
\end{itemize}
Combining the above facts, we can conclude that the optimal value of (P3) is equal to that of (P4). Denote this value as $W_{0}$. According to Theorem \ref{Thm:lower-bound}, $W - W_{0}$ is an lower bound of the clearance time of the index coding problem $\mathcal{G}$. On the other hand, $W - W_{0}$ is the clearance time achieved by the scalar cyclic code corresponding to the solution of (P4), or equivalently (P2). Hence, we can conclude that the cyclic code given by (P2) is the optimal index code.
\end{IEEEproof}


\subsection{Optimality of Cyclic Codes in the Unicast-Uniprior Index Coding Problem}
In this subsection, we consider the unicast-uniprior index coding problem where each packet is demanded by one single user and can be contained as side information by one single user. The problem is motivated by the broadcast relay problem \cite{NeelyTehraniZhang12} where multiple users exchange their individual data through a broadcast relay. 

A strong corollary of Theorem \ref{Thm:cyclic_code_planar_graph} on the unicast-uniprior index coding problem is presented as below. This corollary is also an enhancement of the conclusion in section III.C of \cite{NeelyTehraniZhang12} where the cyclic code is proven to be the optimal index code in the unicast-uniprior index coding problem with less than or equal to $3$ users. 

\begin{Cor}
If the number of users in the unicast-uniprior index coding problem is less than or equal to $4$, then cyclic codes are optimal.
\end{Cor}
\begin{IEEEproof}
Let  $\mathcal{G} = (\mathcal{U},\mathcal{P}, \mathcal{A},\mathcal{W}_{\mathcal{P}})$ be a unicast-uniprior index coding problem where each packet vertex has one single outgoing link and one single incoming link and $|\mathcal{U}| \leq 4$. Let the underlying undirected graph of $\mathcal{G}$ be $U(\mathcal{G})$. The degree of a vertex in an undirected graph is defined as the number of its adjacent edges.  At most $4$ vertices in $U(\mathcal{G})$ can have a degree larger than $2$. That is because each packet vertex must have a degree of $2$ and only a user vertex can have a degree larger than $2$. 

By Theorem \ref{Thm:Wagner1937}, if $U(\mathcal{G})$ is nonplanar, there must exist a subgraph of $U(\mathcal{G})$ which can be converted to either $K_{5}$ or $K_{3,3}$ after several contracting operations.  Note that $K_{5}$ has $5$ nodes with identical degree of $4$ and $K_{3,3}$ has $6$ nodes with identical degree of $3$. Also note that no matter a user-to-packet edge or a packet-to-user edge in $U(\mathcal{G})$ is contracted, one user vertex and one packet vertex are replaced by one new vertex whose degree is equal to the degree of the user vertex. As a result, contracting operations performed over $U(\mathcal{G})$ can not generate new nodes with degree larger than $2$. Thus, there doesn't exist a subgraph of $U(\mathcal{G})$ which has $K_{5}$ or $K_{3,3}$ as minor.  So graph $U(\mathcal{G})$ must be planar. By Theorem \ref{Thm:cyclic_code_planar_graph}, the cyclic code is optimal in $\mathcal{G}$.
\end{IEEEproof}

\section{Partial Clique Codes: A Duality Perspective}\label{section:partial-clique-codes}
Section \ref{section:cyclic-code-duality} shows the inherent duality between the tightest lower bound given by Theorem \ref{Thm:lower-bound} and the optimal cyclic code.  In fact, this is not an isolated case. In this section, a different code structure involving \emph{partial clique codes} is considered. Partial clique codes are more sophisticated but often lead to  performance improvements over cyclic codes. It is shown that the problem of finding the optimal partial clique code is the dual problem of another LP relaxation of (P1). This observation suggests that one could possibly design a good code for the index coding problem by exploring LP relaxations of (P1) and studying their dual problems. 

\subsection{Partial Clique Codes}

Let $\mathcal{P}_0 \subseteq \mathcal{P}$ be a subset of $k (1\leq k\leq M)$ packet vertices and $\displaystyle \mathcal{N}_{\text{out}}(\mathcal{P}_0) =  \bigcup_{p\in \mathcal{P}_0} \mathcal{N}_{\text{out}}(p)$ be the outgoing neighborhood of $p_m$, i.e., the subset of users who demanded packets in $\mathcal{P}_0$.  If each user in $\mathcal{N}_{\text{out}}(P_0)$ has at least $d   (0\leq d \leq k-1)$ packet vertices in $\mathcal{P}_0$ as side information, then the subgraph of $\mathcal{G}$ induced by $\mathcal{P}_0$ and $\mathcal{N}_{\text{out}}(\mathcal{P}_0)$ is a $(k, d)$-partial clique.  A $(k,d)$-partial clique where the weight of each packet vertex is identically $1$ can be cleared with $k-d$ transmissions using $k-d$ independent linear combinations of the packets (such as using Reed-Solomon erasure  codes in \cite{Birk98INFOCOM} or random codes in \cite{Ho06IT}). For example, the digraph $\mathcal{G}$ in Figure \ref{fig:bipartite_graph_example} itself is a $(3,1)$-partial clique. If the weight of each packet vertex is identically one, then this graph can be cleared by transmitting $2$ linear combinations in the form $Z = \alpha_1 p_1 + \alpha_2 p_2 + \alpha_3 p_3$ where $\alpha_i$'s are taken from a finite filed $\mathbb{F}$. If the finite field $\mathbb{F}$ is large enough, we are able to find $2$ linear combinations such that the $2$ linear combinations together with any one in   $p_1$, $p_2$ and $p_3$ are linearly independent. Thus, each user $u_i, i=1,2,3$ can decode $p_i$ by solving a system of $3$ linear equations. 

The linear index code of $\mathcal{G}$ is said to be a partial clique code if it uses a sequence of coding actions that involve only partial clique coding actions. Note that the subgraph induced by a single packet vertex and the user vertex demanding it is by definition a $(1,0)$-partial clique. Let $\mathcal{T}_{k,d}, k=1,\ldots,M, d=0,\ldots,k-1$ be the set of all $(k,d)$-partial cliques in $\mathcal{G}$, then the problem of finding the optimal scalar partial clique code can be formulated as an ILP problem as below:
\begin{align*}
\text{(P5)}
\begin{aligned}
\underset{y_{T_{k,d}}, \forall T_{k,d}\in \mathcal{T}_{k,d},k=1,\ldots,M, d=0,\ldots,k-1}{\text{min}} \quad &\sum_{k=1}^M \sum_{d=0}^{k-1} \sum_{T_{k,d} \in \mathcal{T}_{k,d}} y_{T_{k,d}} (k-d) \\
\text{s.t.} \quad \quad &\sum_{k=1}^M \sum_{d=0}^{k-1} \sum_{T_{k,d} \in \mathcal{T}_{k,d}} y_{T_{k,d}}  \mathbf{1}_{\{p_m \in T_{k,d}\}} \geq w_{p_m},\quad m=1,\ldots, M\\
			&y_{T_{k,d}} \text{ non-negative integral},\quad \forall T_{k,d}\in \mathcal{T}_{k,d},k=1,\ldots,M, d=0,\ldots,k-1
\end{aligned}
\end{align*}
where $y_{T_{k,d}}$ is the number of partial clique codes over each partial clique $T_{k,d}, \forall T_{k,d}\in \mathcal{T}_{k,d}, , k=1,\ldots,M, d=0,\ldots,k-1$, objective function $\sum_{k=1}^M \sum_{d=0}^{k-1} \sum_{T_{k,d} \in \mathcal{T}_{k,d}} y_{T_{k,d}} (k-d)$ is the total number of transmissions, and $\sum_{k=1}^M \sum_{d=0}^{k-1} \sum_{T_{k,d} \in \mathcal{T}_{k,d}} y_{T_{k,d}}  \mathbf{1}_{\{p_m \in T_{k,d}\}} \geq w_{p_m}$ is the constraint that all the $w_{p_m}$ packets represented by packet vertex $p_m$ are cleared by partial cliques involving it.

The problem of finding the optimal vector partial clique code can be formulated as a linear programing problem as below:
\begin{align*}
\text{(P5$^\prime$)}
\begin{aligned}
\underset{y_{T_{k,d}}, \forall T_{k,d}\in \mathcal{T}_{k,d},k=1,\ldots,M, d=0,\ldots,k-1}{\text{min}} \quad &\sum_{k=1}^M \sum_{d=0}^{k-1} \sum_{T_{k,d} \in \mathcal{T}_{k,d}} y_{T_{k,d}} (k-d) \\
\text{s.t.} \quad \quad &\sum_{k=1}^M \sum_{d=0}^{k-1} \sum_{T_{k,d} \in \mathcal{T}_{k,d}} y_{T_{k,d}}  \mathbf{1}_{\{p_m \in T_{k,d}\}} \geq w_{p_m},\quad m=1,\ldots, M\\
&y_{T_{k,d}}\geq0,\quad \forall T_{k,d}\in \mathcal{T}_{k,d},k=1,\ldots,M, d=0,\ldots,k-1
\end{aligned}
\end{align*}
Similar to cyclic codes, (P5$^\prime$) is the LP relaxation of (P5).

The structure of partial clique codes is much more sophisticated than that of cyclic codes. Typically, partial clique codes have to be implemented over a large enough finite field while cyclic codes can always be implemented over the binary field. On the other hand, the performance of partial clique codes in general is better (no worse) than that of cyclic codes. This is summarized in the following lemma.

\begin{Lem}\label{Lem:partial_clique_code_better_than_cyclic_code}
In any (unicast) index coding problem, the optimal clearance time attained by  scalar cyclic codes  is no less than that attained by scalar partial clique codes. Similarly, the optimal clearance time attained by vector cyclic codes  is no less than that attained by vector partial clique codes.
\end{Lem}
\begin{IEEEproof}
This lemma is proven for scalar codes. However, all the arguments can be carried over to vector codes after each packet is divided into subpackets. Recall that in any $K$-cycle, each user vertex has at least one packet vertex as side information. So each $K$-cycle code can be equivalently replaced by a $(K,1)$-partial clique code. This uses partial clique coding to achieve the same clearance time.  Thus, the best partial clique coding strategy achieves a clearance time that is less than or equal to that of the best cyclic coding strategy.
\end{IEEEproof}

\begin{figure}[htbp]
   \centering
   \includegraphics[width=4in]{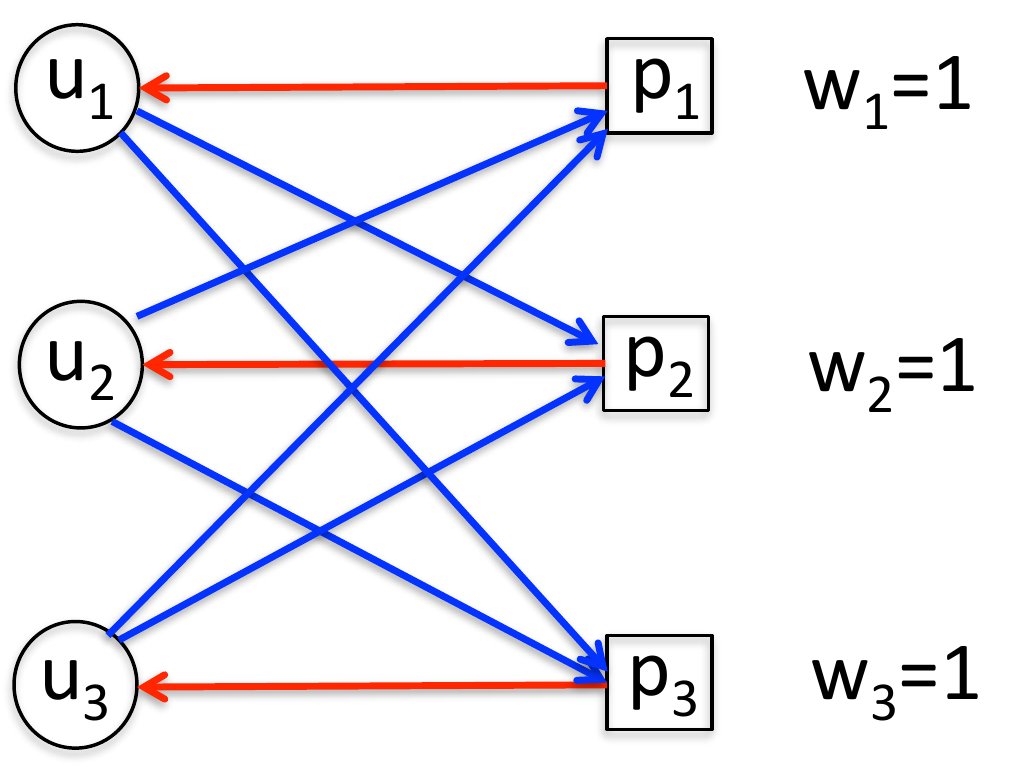} 
   \caption{An example with $3$ users and $3$ packets where the partial clique code is strictly better than the cyclic code.}
   \label{fig:partial_clique_strictly_better}
\end{figure}

Figure \ref{fig:partial_clique_strictly_better} shows an example of the index coding problem with $3$ users and $3$ packets. The bipartite digraph of this problem is not planar. (In fact, this example is the only unicast index coding problem with $3$ users and $3$ packets for which the bipartite digraph is non-planar.) It can be verified that the optimal scalar cyclic code can clear this problem with $2$ transmissions. On the other hand, the bipartite digraph itself is a $(3,2)$-partial clique and hence the scalar partial clique code can clear it with one single transmission. The scalar partial clique code simply transmits $Z= p_1 +p_2 +p_3$. In this simple example, the scalar partial clique code is strictly better than the scalar cyclic code. However, the following theorem shows that partial clique codes have no performance advantage over cyclic codes in the unicast-uniprior index coding problem.

\begin{Thm}\label{Thm:unicast-unipior-cyclic-eq-clique}
In any unicast-uniprior index coding problem, the optimal clearance time attained by scalar cyclic codes is equal to that attained by scalar partial cliques.  Similarly, the optimal clearance time attained by vector cyclic codes is equal to that attained by vector partial cliques. 
\end{Thm}
\begin{IEEEproof}
This theorem is proven for scalar codes. However, all the arguments can be carried over to vector codes after each packet is divided into subpackets. 
\begin{itemize}
\item Claim 1: The optimal clearance time attained by cyclic codes is larger than or equal to that attained by partial clique codes. This is Lemma \ref{Lem:partial_clique_code_better_than_cyclic_code}.
\item Claim 2: The optimal clearance time attained by cyclic codes is less than or equal to that attained by partial clique codes. For any partial clique $T_{k,d}$ ($d\geq 1$) utilized in the optimal partial clique code, $k$ packets are cleared with $k-d$ transmissions. By definition of partial cliques, each user vertex in this $T_{k,d}$ has at least $d$ arcs outgoing to packet vertices in it. So we are able to find a cycle in it. To find a cycle, we start at any vertex, traverse a path from vertex to vertex using any outgoing link and discover a cycle when we revisit a vertex. Denote this cycle as $C_1$ and delete all the packet vertices and the associated outgoing and incoming arcs from $T_{k,d}$. Note that each packet vertex has one single outgoing and one single incoming arc in a unicast-uniprior index coding problem.  Hence, no two packet vertices in $C_1$ share the same outgoing neighbor or  incoming neighbor. So after the deletion of the packet vertices and the associated outgoing and incoming arcs, the number of outgoing arcs of the user vertices involved in $C_1$ decreases by one while the number of outgoing arcs of the user vertices not involved in $C_1$ does not change. So in the remaining part of this $T_{k,d}$, each user vertex has at least $d-1$ outgoing arcs. Repeat the above process again and again. In the end, we have $d$ cycles and no two cycles share the same packet vertex. So by performing a cycle code over each cycle $C_i, i=1,\ldots,d$, we can save $d$ transmissions in total. Hence, this $T_{k,d}$ can be cleared with $k-d$ transmissions by applying cyclic codes. As a result, cyclic codes are no worse than partial clique codes in the unicast-uniprior index coding problem. 
\end{itemize}
\end{IEEEproof}

\subsection{Duality Between Information Theoretical Lower Bounds and Partial Clique Codes}

Define an ILP problem as below:
\begin{align*}
\text{(P6)}
\begin{aligned}
\underset{x_m, m=1,\ldots,M}{\text{max}} \quad &\sum_{m=1}^M x_m w_{p_m}\\
\text{s.t.} \quad \quad &\sum_{m=1}^M x_m \mathbf{1}_{\{p_m \in T_{k,d}\}} \leq k-d, \quad \forall T_{k,d}\in \mathcal{T}_{k,d}, k=1,\ldots,M, d=0,\ldots,k-1\\
			&x_m \in \{0,1\},\quad m=1,\ldots, M
\end{aligned}
\end{align*}
The physical meaning of (P6) is to find the maximum packet weighted subgraph of $\mathcal{G}$ formed by packet vertex deletions such that  at least $d$ packet vertices are deleted in each $(k,d)$ partial clique. 

\begin{Lem}\label{Lem:LP_duality_partial_clique_code}
(P5$^\prime$)  and the LP relaxation of (P6) are a primal-dual linear programming pair.
\end{Lem}
\begin{IEEEproof}
The Lagrangian function of (P5$^\prime$) can be written as 
\begin{align*}
L(y_{T_{k,d}}, \lambda_m, \mu_{T_{k,d}}) &= \sum_{k=1}^M \sum_{d=0}^{k-1} \sum_{T_{k,d} \in \mathcal{T}_{k,d}} y_{T_{k,d}} (k-d)  + \sum_{m=1}^M \lambda_m \big[ w_{p_m}  - \sum_{k=1}^M \sum_{d=0}^{k-1} \sum_{T_{k,d}\in \mathcal{T}_{k,d}} y_{T_{k,d}} \mathbf{1}_{\{p_m \in T_{k,d}\}} \big]  \\ & \quad - \sum_{k=1}^M \sum_{d=0}^{k-1} \sum_{T_{k,d}\in \mathcal{T}_{k,d}} \mu_{T_{k,d}} y_{T_{k,d}} \\
&= \sum_{m=1}^M \lambda_m w_{p_m} + \sum_{k=1}^M \sum_{d=0}^{k-1} \sum_{T_{k,d}\in\mathcal{T}_{k,d}} y_{T_{k,d}} \big[ (k-d) - \sum_{m=1}^M \lambda_m \mathbf{1}_{\{p_m \in T_{k,d}\}} - \mu_{T_{k,d}}\big] 
\end{align*}
where $\lambda_m \geq 0, m=1,\ldots,M$ and $\mu_{T_{k,d}}\geq 0, \forall T_{k,d} \in \mathcal{T}_{k,d} , k=1,\ldots,M, d=0,\ldots,k-1$. The dual problem of (P5$^\prime$) is defined as:
\begin{align*}
\underset{\begin{subarray}{c}\lambda_m \geq 0, m=1,\ldots,M;\\\mu_{T_{k,d}}\geq 0, \forall T_{k,d}\in \mathcal{T}_{k,d},k=1,\ldots,M, d=0,\ldots,k-1;\end{subarray}}{\max}  \underset{\begin{subarray}{c} y_{T_{k,d}}\in \mathbb{R}, k=2,\ldots,M,d=0,\ldots,k-1, \forall T_{k,d} \in\mathcal{T}_{k,d}\end{subarray}} {\min}  L(y_{T_{k,d}}, \lambda_m, \mu_{T_{k,d}})\end{align*}
Note that, 
\begin{align*}
\underset{y_{T_{k,d}}\in \mathbb{R}, k=2,\ldots,M,d=0,\ldots,k-1, \forall T_{k,d} \in\mathcal{T}_{k,d}} {\min}  L(y_{T_{k,d}}, \lambda_m, \mu_{T_{k,d}}) = \left\{ \begin{array}{cc}
\sum_{m=1}^M \lambda_m w_{p_m} &  \begin{subarray}{l} (k-d) - \sum_{m=1}^M \lambda_m \mathbf{1}_{\{p_m \in T_{k,d}\}} - \mu_{T_{k,d}} =0 \\ \forall T_{k,d}\in \mathcal{T}_{k,d},k=1,\ldots,M, d=0,\ldots,k-1 \end{subarray}\\-\infty & \text{otherwise}\end{array}\right.
\end{align*}
Then, the dual problem of (P5$^\prime$) can be written as,
\begin{align*}
\begin{aligned}
\underset{\begin{subarray}{c}\lambda_m, m=1,\ldots,M;\\\mu_{T_{k,d}}, \forall T_{k,d}\in \mathcal{T}_{k,d},k=1,\ldots,M, d=0,\ldots,k-1;\end{subarray}}{\text{max}} \quad &\sum_{m=1}^M \lambda_m w_{p_m}\\
\text{s.t.} \quad \quad &(k-d) - \sum_{m=1}^M \lambda_m \mathbf{1}_{\{p_m \in T_{k,d}\}} - \mu_{T_{k,d}} =0, \quad \begin{subarray}{c} \forall T_{k,d} \in \mathcal{T}_{k,d} \\k=1,\ldots,M, d=0,\ldots,k-1,\end{subarray}\\
			&\lambda_m \geq 0, \quad m=1,\ldots,M\\
			&\mu_{T_{k_d}} \geq 0, \quad \forall T_{k,d}\in \mathcal{T}_{k,d},k=1,\ldots,M, d=0,\ldots,k-1\\
\end{aligned}
\end{align*}
Eliminating variables $\mu_{T_{k_d}}, \forall T_{k,d}\in \mathcal{T}_{k,d},k=1,\ldots,M, d=0,\ldots,k-1$, we obtain
\begin{align*}
\begin{aligned}
\underset{\begin{subarray}{c}\lambda_m, m=1,\ldots,M\end{subarray}}{\text{max}} \quad &\sum_{m=1}^M \lambda_m w_{p_m}\\
\text{s.t.} \quad \quad &\sum_{m=1}^M \lambda_m \mathbf{1}_{\{p_m \in T_{k,d}\}}  \leq (k-d), \quad \begin{subarray}{c} \\ \forall T_{k,d} \in \mathcal{T}_{k,d}, \forall  k=1,\ldots,M, d=0,\ldots,k-1\end{subarray}\\
			&\lambda_m \geq 0, \quad m=1,\ldots,M\\
\end{aligned}
\end{align*}
Now consider all the $M$ packet vertices, i.e., all $T_{1,0} \in \mathcal{T}_{1,0}$. The corresponding constraints $\sum_{m=1}^M \lambda_m \mathbf{1}_{\{p_m \in T_{k,d}\}}  \leq (k-d), \forall T_{1,0} \in \mathcal{T}_{1,0}$ can be simplified as $\lambda_m \leq 1, m=1,\ldots,M$. Hence, the above linear programming problem is the LP relaxation of (P6).
\end{IEEEproof}

Integer program (P6) seems quite different from (P1) and it seems that there exists no duality between the optimal partial clique code and the tightest lower bound.  However, the following lemma shows that (P1) and (P6) are two equivalent problems.
\begin{Lem}\label{Lem:equivalet_lower_bound}
For any unicast index coding problem $\mathcal{G}$,  (P1) and (P6) are two equivalent problems.
\end{Lem}
\begin{IEEEproof}
Note that the objective function in (P1) is the same as that in (P6). To prove problem (P1) and (P6) are equivalent, we show that $x_m \in \{0,1\}, m=1,\ldots, M$ is feasible to (P1) if and only if it is feasible to (P6).
\begin{itemize}
\item ``if" part: Assume $x_m \in \{0,1\}, m=1,\ldots, M$ is feasible to (P6). For any cycle $C_i, \forall C_{i}\in \mathcal{C}_i, i=2,\ldots,L$ involving $i$ packet vertices in $\mathcal{G}$, let us consider the partial clique $T_{i,d}$ formed by the $i$ packet vertices and $i$ user vertices in this $i$-cycle. By the definition of a cycle, each user vertex has at least one packet vertex among these $i$ packet vertices as side information. So $d\geq 1$. Since $x_m \in \{0,1\}, m=1,\ldots, M$ satisfies the inequality constraints in (P6), at least $d$ packet vertices among these $i$ packet vertices are deleted. So cycle $C_{i}$ can not be complete. Hence,  $x_m,m=1,\ldots, M$ yields a acyclic subgraph of $\mathcal{G}$.
\item ``only if" part: Assume $x_m \in \{0,1\}, m=1,\ldots, M$ is feasible to (P1). For any partial clique $T_{k,d}$, if $d=0$, then constraint $\sum_{m=1}^M x_m \mathbf{1}_{\{p_m \in T_{k,d}\}} \leq k-d$ is trivially satisfied. Without loss of generality, assume $1\leq d \leq k-1$. Then, in this partial clique $T_{k,d}$, each user vertex has at least $d$ outgoing arcs.  So we can find a cycle in this partial clique. (To find a cycle, we start at any vertex, traverse a path from vertex to vertex using any outgoing link and discover a cycle when we revisit a vertex.) Since $x_m \in \{0,1\}, m=1,\ldots, M$ is feasible to (P1), at least one packet vertex in this cycle is deleted. Assume $d_1$ packet vertices are deleted. These deleted packet vertices are also vertices in partial clique $T_{k,d}$. If $d_1 =d$, then the constraint over $T_{k,d}$ is satisfied. If $d_1 < d$, then we continue to consider the remaining part of $T_{k,d}$ after deleting these $d_1$ packet vertices. In the remaining part, each user vertex has at least $d-d_1$ outgoing arcs. A similar argument as above shows that we are still able to find a new cycle in the remaining part and at least one packet vertex in the cycle is deleted.  Assume $d_2$ packet vertices in the new cycle are deleted. If $d_1 + d_2 < d$, we can repeat this process again until at least $d$ packet vertices are shown to be deleted. That is to say, constraint $\sum_{m=1}^M x_m \mathbf{1}_{\{p_m \in T_{k,d}\}} \leq k-d$ over all $T_{k,d}$ is satisfied. Hence, $x_m \in \{0,1\}, m=1,\ldots, M$ satisfies the constraints of (P6).
\end{itemize}
\end{IEEEproof}

The above lemma indicates that (P6) is another representation of (P1). However, this new representation is non-trivial. The LP relaxations of (P6) and (P1) correspond to partial clique codes and cyclic codes, respectively. Lemma \ref{Lem:partial_clique_code_better_than_cyclic_code} demonstrates that codes associated with (P6) in general have better performance than codes associated with (P1).  

\subsection{Discussions}

\begin{figure}
   \centering
   \includegraphics[width=7.5in]{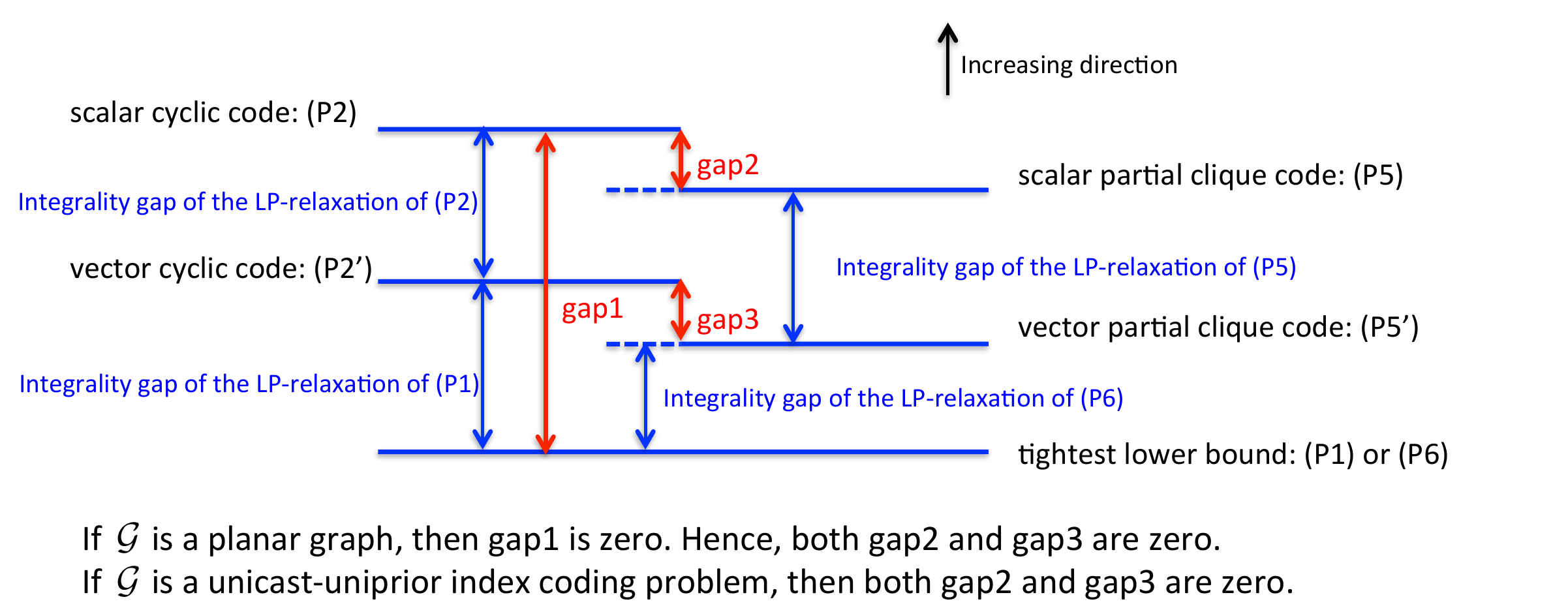} 
   \caption{The relations between various problems in this paper.}
   \label{fig:comparison}
\end{figure}

In the subject of integer linear programming, (P1) and (P6) are considered as two different representations of the same integer linear program. However, different representations of an integer linear program can yield different LP relaxations. The optimal values, or equivalently the integrality gaps, of different LP relaxations can be quite different. In section \ref{section:cyclic-code-duality} and this section, we show that the LP relaxation of (P1) is the (dual) problem of finding the optimal vector cyclic code and the LP relaxation of (P6) is the (dual) problem of finding the optimal vector partial clique code. The performance of partial clique codes is no worse than that of cyclic codes. This is because the integrality gap of the LP relaxation of (P6) is no larger than that of the LP relaxation of (P1). The relations between various problems in this paper are illustrated in Figure \ref{fig:comparison}.

Since there are various techniques on how to obtain tight LP relaxations of an integer linear program\cite{book_IntegerCombinatorialOpt},  a potential approach to design good code structures for the index coding problem is to explore different representations of (P1) for which the LP relaxations have small integrality gaps and study their dual problems. If the dual problem of any LP relaxation can be interpreted as a code structure, then this is a good code for the index coding problem.

\section{Conclusion}
This paper studies index coding from a perspective of optimization and duality.  It illustrates the inherent duality between the information theoretical lower bound, defined by the maximum acyclic subgraph, and the optimal cyclic codes and partial clique codes. The performance of both codes is bounded by the respective integrality gap of two different LP relaxations of the tightest lower bound problem. In the special case when the index coding problem has a planar digraph representation, the integrality gap associated with cyclic coding is shown to be zero. So the exact optimality is achieved by cyclic coding. For general (non-planar) problems, the LP-relaxation associated with partial clique coding provides an integrality gap that is no worse, and often better, than the previous gap. This ensures that partial clique coding is no worse, and often better, than cyclic coding. These results provide new insight into the index coding problem and suggest that good codes can be found by exploring the LP relaxations of the tightest lower bound problem.

\bibliographystyle{IEEEtran}
\bibliography{IEEEfull,mybibfile}

\end{document}